\documentclass[twocolumn,epjc3]{svjour3}
\usepackage{amsmath}
\usepackage{amssymb}
\usepackage[english]{babel}
\usepackage[utf8]{inputenc}
\usepackage[font=small, format=hang, labelfont={bf}]{caption}
\usepackage{graphicx}
\usepackage{epstopdf}
\usepackage{float}
\graphicspath{{Immagini/}}
\usepackage{longtable}
\usepackage[a4paper,top=2cm,bottom=1.8cm,left=2.5cm,right=2.5cm]{geometry}
\usepackage{psfrag}
\usepackage{subfig}
\usepackage[T1]{fontenc}
\usepackage{textcomp}
\usepackage[
citecolor=black,%
filecolor=black,%
linkcolor=black,%
urlcolor=blue,%
colorlinks=true]{hyperref}
\usepackage[absolute,overlay]{textpos}
\usepackage{graphicx}
\usepackage{multirow}
\usepackage{adjustbox}
\usepackage{xspace}

\usepackage{lineno}

\newcommand{\myl}{Mylar\textregistered\xspace}
\title{\boldmath Enhancing the light yield of He:CF$_4$ based gaseous detector}

\author{Fernando Domingues Amaro\thanksref{1} \and Rita Antonietti\thanksref{2,3} \and Elisabetta Baracchini\thanksref{4,5} \and Luigi Benussi\thanksref{6} \and Stefano Bianco\thanksref{6} \and Roberto Campagnola\thanksref{6} \and Cesidio Capoccia\thanksref{6} \and Michele Caponero\thanksref{6,7} \and Danilo Santos Cardoso\thanksref{8} \and Luan Gomes Mattosinhos de Carvalho\thanksref{9} \and Gianluca Cavoto\thanksref{10,11} \and Igor Abritta Costa\thanksref{6} \and Antonio Croce\thanksref{6} \and Emiliano Dan\'e\thanksref{6} \and Giorgio Dho\thanksref{4,6,e}
\and Flaminia Di Giambattista\thanksref{4,5} \and Emanuele Di Marco\thanksref{11} \and Melba D'Astolfo\thanksref{4,5} \and Giulia D'Imperio\thanksref{11} \and Davide Fiorina\thanksref{4,5} \and Francesco Iacoangeli\thanksref{11} \and Zahoor ul Islam\thanksref{4,5} \and Herman Pessoa Lima J\`unior\thanksref{4,8} \and Ernesto Kemp\thanksref{12} \and Giovanni Maccarrone\thanksref{6} \and Rui Daniel Passos Mano\thanksref{1} \and Robert Renz Marcelo Gregorio\thanksref{13} \and David Jos\'e Gaspar Marques\thanksref{4,5} \and Giovanni Mazzitelli\thanksref{6} \and Alasdair Gregor McLean\thanksref{13} \and Andrea Messina\thanksref{10,11} \and Pietro Meloni\thanksref{2,3} \and Cristina	Maria Bernardes	Monteiro\thanksref{1} \and Rafael Antunes Nobrega\thanksref{9} \and Igor Fonseca Pains\thanksref{9} \and Emiliano Paoletti\thanksref{6} \and Luciano Passamonti\thanksref{6} \and Fabrizio Petrucci\thanksref{2,3} \and Stefano Piacentini\thanksref{4,5} \and Davide Piccolo\thanksref{6} \and Daniele Pierluigi\thanksref{6} \and Davide Pinci\thanksref{11} \and Atul Prajapati\thanksref{4,5} \and Francesco Renga\thanksref{11} \and Rita	Joana da Cruz	Roque\thanksref{1} \and Filippo Rosatelli\thanksref{6} \and Alessandro Russo\thanksref{6} \and Joaquim	Marques Ferreira dos Santos\thanksref{1} \and Giovanna Saviano\thanksref{6,14} \and Pedro Alberto Oliveira Costa Silva\thanksref{1} \and Neil	John Curwen	Spooner\thanksref{13} \and Roberto Tesauro\thanksref{6} \and Sandro Tomassini\thanksref{6} \and Samuele Torelli\thanksref{4,5} 
}
\thankstext{e}{\emph{Coresponding author:} giorgio.dho@lnf.infn.it}

\institute{LIBPhys; Department of Physics; University of Coimbra; 3004-516 Coimbra; Portugal\label{1} \and Dipartimento di Matematica e Fisica; Universit\`a Roma TRE; 00146; Roma; Italy\label{2} \and Istituto Nazionale di Fisica Nucleare; Sezione di Roma Tre; 00146; Rome; Italy\label{3} \and Gran Sasso Science Institute; 67100; L'Aquila; Italy\label{4} \and Istituto Nazionale di Fisica Nucleare; Laboratori Nazionali del Gran Sasso; 67100; Assergi; Italy\label{5} \and Istituto Nazionale di Fisica Nucleare; Laboratori Nazionali di Frascati; 00044; Frascati; Italy\label{6} \and ENEA Centro Ricerche Frascati; 00044; Frascati; Italy\label{7}  \and Centro Brasileiro de Pesquisas F\'isicas; Rio de Janeiro 22290-180; RJ; Brazil\label{8} \and Universidade Federal de Juiz de Fora; Faculdade de Engenharia; 36036-900; Juiz de Fora; MG; Brasil\label{9} \and Dipartimento di Fisica; Sapienza Universit\`a di Roma; 00185; Roma; Italy\label{10} \and Istituto Nazionale di Fisica Nucleare; Sezione di Roma; 00185; Rome; Italy\label{11}  \and Universidade Estadual de Campinas (UNICAMP); Campinas 13083-859; SP; Brazil\label{12}  \and Department of Physics and Astronomy; University of Sheffield; Sheffield; S3 7RH; UK\label{13} \and Dipartimento di Ingegneria Chimica; Materiali e Ambiente; Sapienza Universit\`a di Roma; 00185; Roma; Italy\label{14}}

\date{Internal v1: 2024-05-13}

\begin{document}
\maketitle

\begin{abstract}
The CYGNO experiment aims to build a large ($\mathcal{O}(10)$ m$^3$) directional detector for rare event searches, such as nuclear recoils (NRs) induced by dark matter (DM), such as weakly interactive massive particles (WIMPs). The detector concept comprises a time projection chamber (TPC), filled with a He:CF$_4$ 60/40 scintillating gas mixture at room temperature and atmospheric pressure, equipped with an amplification stage made of a stack of three gas electron multipliers (GEMs) which are coupled to an optical readout. The latter consists in scientific CMOS (sCMOS) cameras and photomultipliers tubes (PMTs). The maximisation of the light yield of the amplification stage plays a major role in the determination of the energy threshold of the experiment. In this paper, we simulate the effect of the addition of a strong electric field below the last GEM plane on the GEM field structure and we experimentally test it by means of a 10$\times$10 cm$^2$ readout area prototype. The experimental measurements analyse stacks of different GEMs and helium concentrations in the gas mixture combined with this extra electric field, studying their performances in terms of light yield, energy resolution and intrinsic diffusion. It is found that the use of this additional electric field permits large light yield increases without degrading intrinsic characteristics of the amplification stage with respect to the regular use of GEMs.

\end{abstract}
\flushbottom

\section{Introduction}
\label{sec:intro}
Currently, the existence of large quantity of non-electromagnetic interacting form of matter in the Universe, referred to as dark matter (DM), is an established and yet puzzling paradigm
\cite{Bertone:2004pz}. Unveiling its nature is one of the frontier studies of modern physics. A possible and well motivated candidate, predicted both by an extension of the Standard Model and by the leading cosmological model ($\Lambda$-CDM), is the weakly interactive massive particle (WIMP), a neutral, stable particle with very low cross section for interaction with standard matter, and a mass that can range between hundreds of MeV/c$^2$ to hundreds of GeV/c$^2$. The measurements of the rotation velocity curves of our Galaxy support the hypothesis of the Standard Halo model, according to which a halo of DM envelopes our Galaxy. Due to the Earth's motion with respect to the centre of the Galaxy, an apparent wind of Dark Matter particles is generated and can be used to experimentally detect them through scattering against regular matter. The effort of the direct detection experiments consists in the exposure of a large mass of sensitive volume in order to be able to detect the very rare occurrences of interactions between WIMPs and nuclei of the target, resulting in nuclear recoils.\\
Theoretically, from these nuclear recoils, it is possible to extract not only the  energetic information, as all the current experiments are capable of, but also their direction and hence the angular and energy distribution of the WIMPs. Determining the incoming direction of the WIMPs can provide a correlation with an astrophysical source that can not be mimicked by any known background    \cite{Mayet_2016zxu}, offering a unique key for a positive identification of a DM signal.\\
The CYGNO experiment \cite{Amaro:2022gub} is following an innovative path for directional DM searches by using a gaseous time projection chamber  (TPC) operated at atmospheric pressure and room temperature, coupled to an optical readout through photomultiplier tubes (PMTs) and scientific CMOS cameras (sCMOS) to measure energy and direction of low energy nuclear recoils. The He:CF$_4$ based gas mixture grants sensitivity to DM masses of O(1) GeV/c$^2$ for both Spin Independent and Dependent coupling, a parameter space still partially uncharted and theoretically well motivated \cite{bib:zurek,bib:petraki,bib:relic}. In CYGNO, a stack of three 50 $\mu$m thick gas electron multipliers (GEMs) is employed as amplification stage, in order to generate electron avalanches that in turn produce light thanks to the scintillating properties of CF$_4$. The ratio of photon to electron produced during electron avalanche in He:CF$_4$ is about 0.1, depending on the gas fractions \cite{bib:GEMoptical,bib:Morozov_2012}, implying a reduction of available signal, which in turn can affect the detection threshold. Moreover, while the optical system coupled to the sCMOS allows to image large areas properly distancing a single detector, it has the drawback of strongly reducing the solid angle covered by sCMOS sensor, hence the amount of light collected. The geometrical acceptance can be as low as 10$^{-4}$ for an imaged area of 25.6 $\times$ 25.6 cm$^2$ as in the LEMOn detector illustrated in Sec. \ref{sec:lemon}. Thus, enhancing the production of photons by the amplification stage is of uttermost importance for the CYGNO experiment in particular and for any gaseous detector exploiting optical readout in general. Increasing the voltage across the GEMs does not solve the problem as one would eventually face breakdown effects in the gas, which disrupts the operation. In addition, larger gain implies larger energy for the avalanche electrons that would further diffuse in the gap between two GEMs, worsening the detector position resolution. 

In \cite{bib:EL_cygno} we demonstrated the possibility to enhance the light yield of He:CF$_4$-based gaseous detector by further accelerating the avalanche electrons after the last GEM with a strong O(10) kV/cm electric induction field. In this paper, we present an additional validation of the results of \cite{bib:EL_cygno} and we extend our studies to larger applied fields, different GEM thicknesses and stacking options and different He:CF$_4$ ratio in the gas mixtures. The paper is organised as following: in Sec. \ref{sec:scintill} we recall the scintillating properties of He:CF$_4$ gas mixtures, in Sec. \ref{sec:maxwell} we present the simulation of the electric fields between the GEMs and the induction gap further supporting our case study, in Sec. \ref{sec:lemon} we illustrate a series of measurements performed with a 7 l active volume detector aimed to further validate the results of \cite{bib:EL_cygno} thanks to a precise evaluation of the charges at play during the light amplification beyond the last GEM, in Sec. \ref{sec:MANGO} we expand the study of such phenomenon to different GEM thicknesses and stacking options also varying the He to CF$_4$ ratio, and in Sec. \ref{sec:disc} we discuss the results.

\section{Scintillating properties of the He:CF$_4$ gas}
\label{sec:scintill}
It is deemed relevant for a better comprehension of the paper to summarise the main characteristics of the scintillating properties of He:CF$_4$ gas mixtures. These properties were studied in details in \cite{bib:Morozov_2012,bib:Kurihara,bib:Margato_2013,bib:Fraga:2003uu}. The light emission spectrum comprises two continua peaked around 290 nm and 620 nm, respectively. The emission in the region centred on the 620 nm peak results from the de-excitation from a Rydberg state of the neutral CF*$_3$ originated from the fragmentation of CF$_4$, with an energy threshold of about 12.5 eV. The dissociative ionisation threshold, on the other hand, is about 15.9 eV. Fig. 5b of Ref. \cite{bib:Kurihara} shows the rate coefficient of the momentum transfer and different excitations, attachment, ionisation, dissociation  for CF$_4$ gas interaction with electrons as a function of the reduced electric field. In particular, the process which refers to the neutral fragmentation responsible for the production of visible light possesses a smaller reduced field threshold than the ionisation ones. This implies that it is theoretically possible to produce light from CF$_4$ without generating charge. Those cross sections are of pure CF$_4$, while the standard CYGNO gas mixture contains large amounts of helium. The cross sections for these mixtures are not found in literature, thus no conclusive assessment on the nature of the light to charge ratio production can be done. Yet, since the primary ionisation energy of He is nearly twice that of CF$_4$, it is fair to assume that the thresholds of ionisation and fragmentation of CF$_4$ do not change significantly from this plot for the gas mixtures under study, even in presence of electric field. 
\section{Maxwell simulation}
\label{sec:maxwell}
\begin{figure}[!t] 
	\centering
	\includegraphics[width=0.49\linewidth]{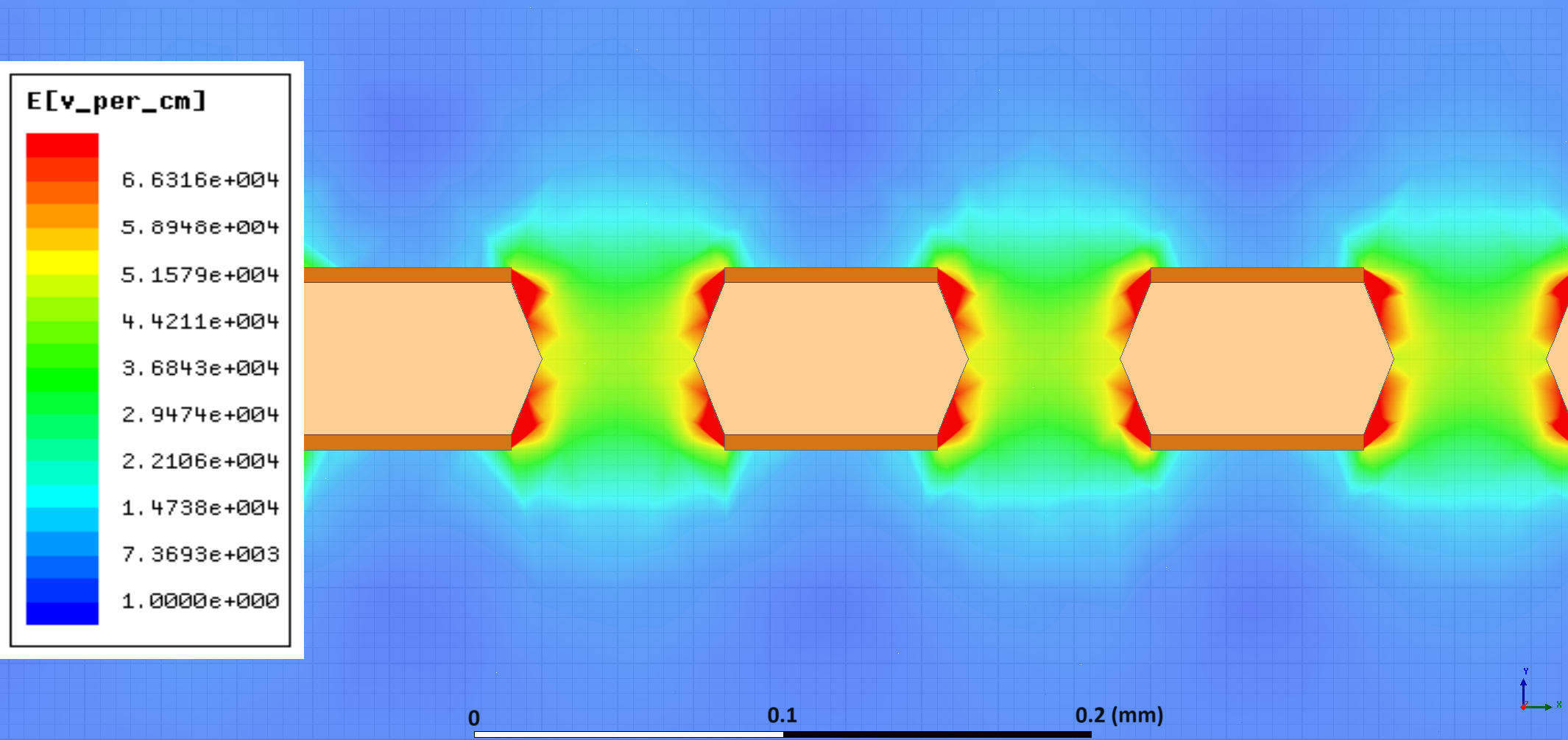}
	\includegraphics[width=0.49\linewidth]{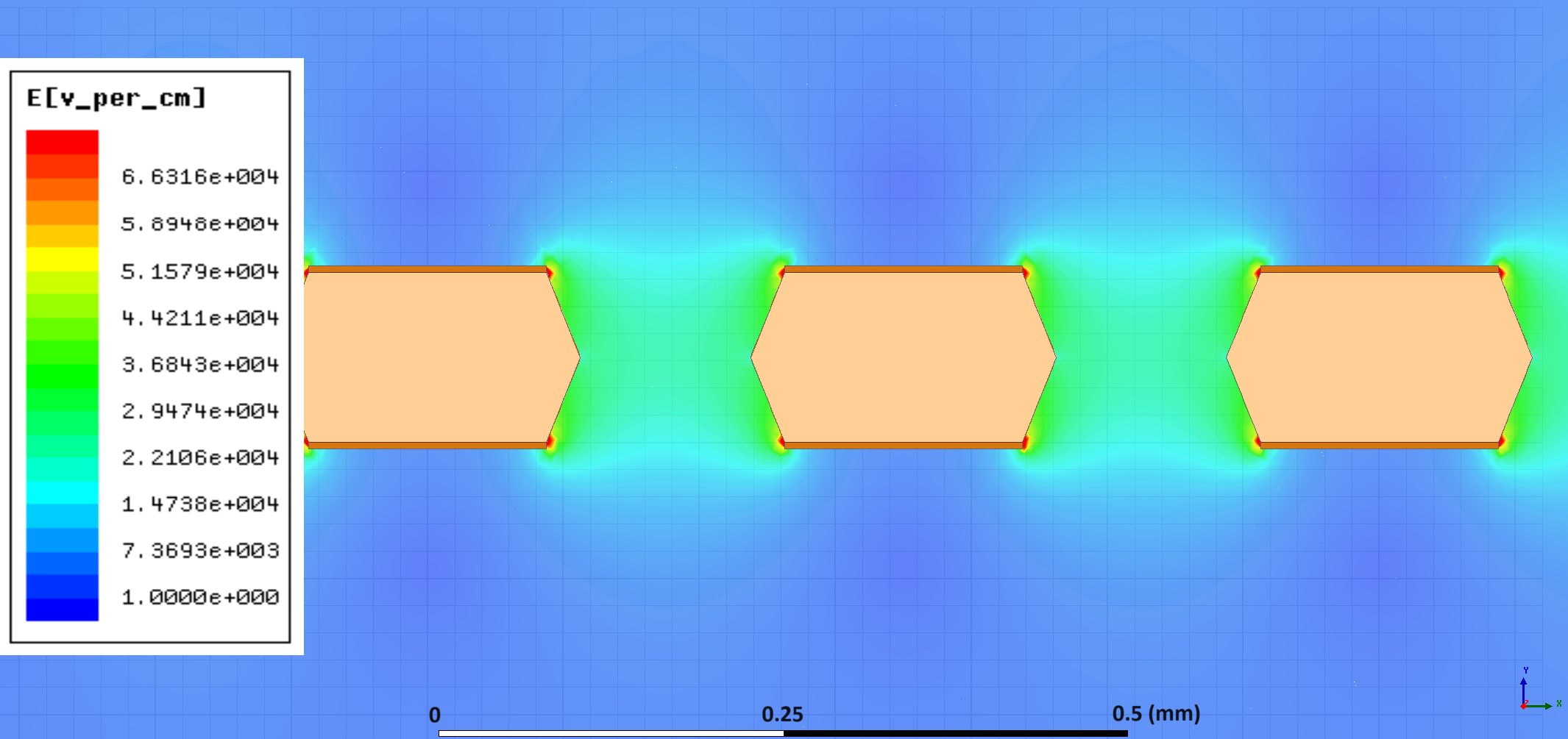}
	\caption{Examples of the 2D electric field maps generated by the Ansys Maxwell program. The vertical axis to the drift direction. The colour scale represents the intensity of the field, with red being the highest one. On the left, the detailed structure of the GEM holes for one thin GEM (50 $\mu$m GEM with 70 $\mu$m radius holes and 140 $\mu$m pitch) with 400 V applied across the GEM, 1000 V applied to a metallic electrode below the GEM and a transfer field of 0 kV/cm above the GEM. On the right, the same for a thick one (125 $\mu$m GEM with 175 $\mu$m radius holes and 350 $\mu$m pitch) with 490 V applied across the GEM, 1000 V applied to the metallic electrode and a transfer field of 0 kV/cm above the GEM.}
	\label{fig:scheme}
\end{figure}
The possibility of enhancing the light yield without substantial production of charge, hence possibly limiting the degradation of energy resolution and diffusion, and the results of Ref. \cite{bib:EL_cygno} support the idea that a strong electric field below the outermost GEM, henceforth called \emph{induction} field or E$_{ind}$, can lead to considerable advantages to the CYGNO optical readout.\\
In order to understand the effect of strong E$_{ind}$ on a GEM-based TPC detector, the electric field characteristics inside, above and below the GEM holes are investigated through simulation. The study is performed with Ansys Maxwell 15\footnote{\url{https://www.ansys.com/products/electronics/ansys-maxwell}}, a commercial software that allows to solve the electromagnetic equations and to obtain the electric field configuration of a specific geometry, among other features. 
\begin{figure}[!t] 
	\centering
	\includegraphics[width=0.49\linewidth]{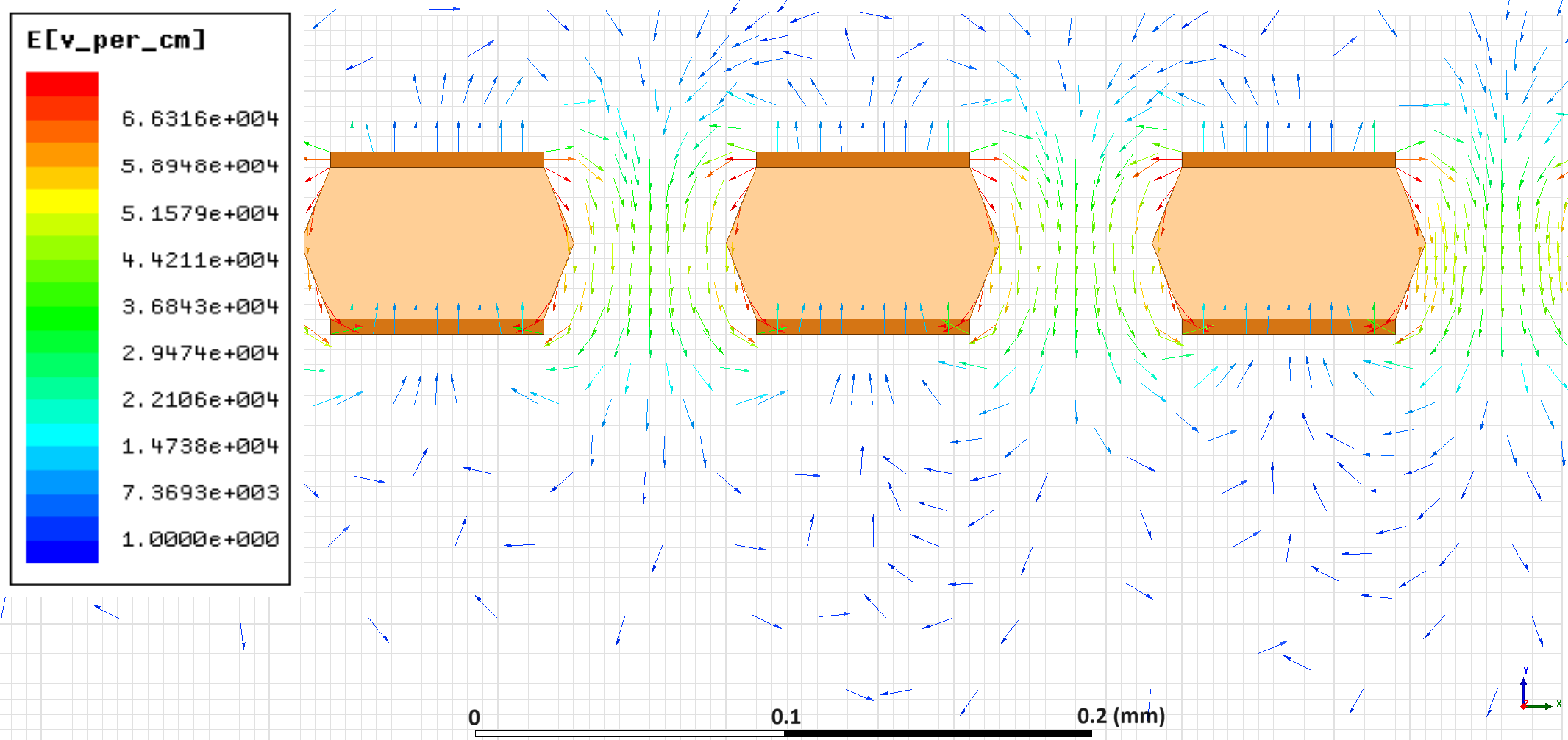}
	\includegraphics[width=0.49\linewidth]{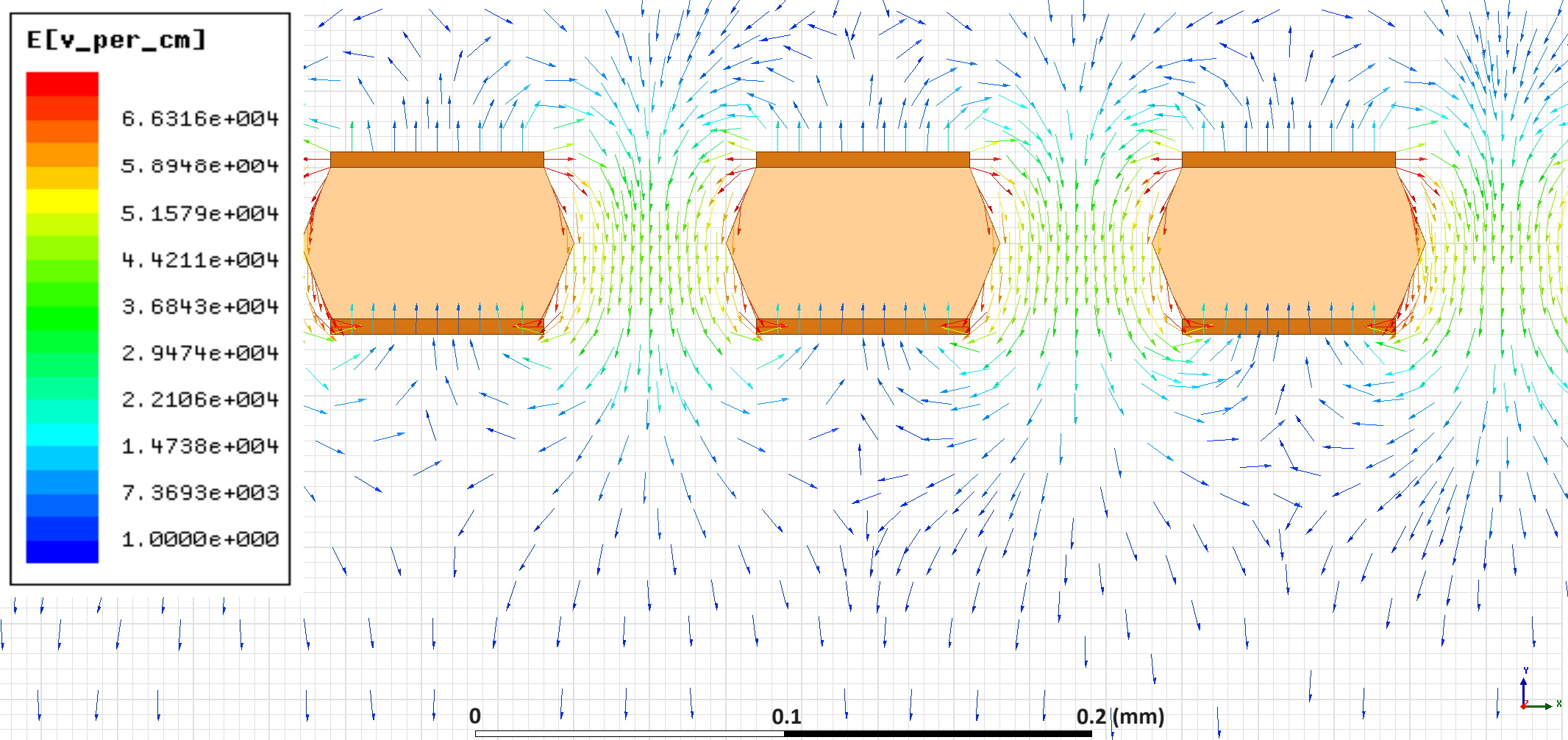}
	\caption{Examples of the 2D electric field vector maps generated by the Ansys Maxwell program. The vertical axis corresponds to the drift direction. The line colour scale represents the intensity of the field, with red being the highest one. On the left, the detailed structure of the GEM holes when no induction field is applied, whilst on the right the same for 1 kV/cm of induction field. It is clearly visible how the field vectors are much more ordered and straight towards the induction gap (bottom of the plot) in the right panel than in the left one, as a result of the induction field addition.}
	\label{fig:flines}
\end{figure}
\begin{figure*}[ht] 
	\centering
	\includegraphics[width=0.9\linewidth]{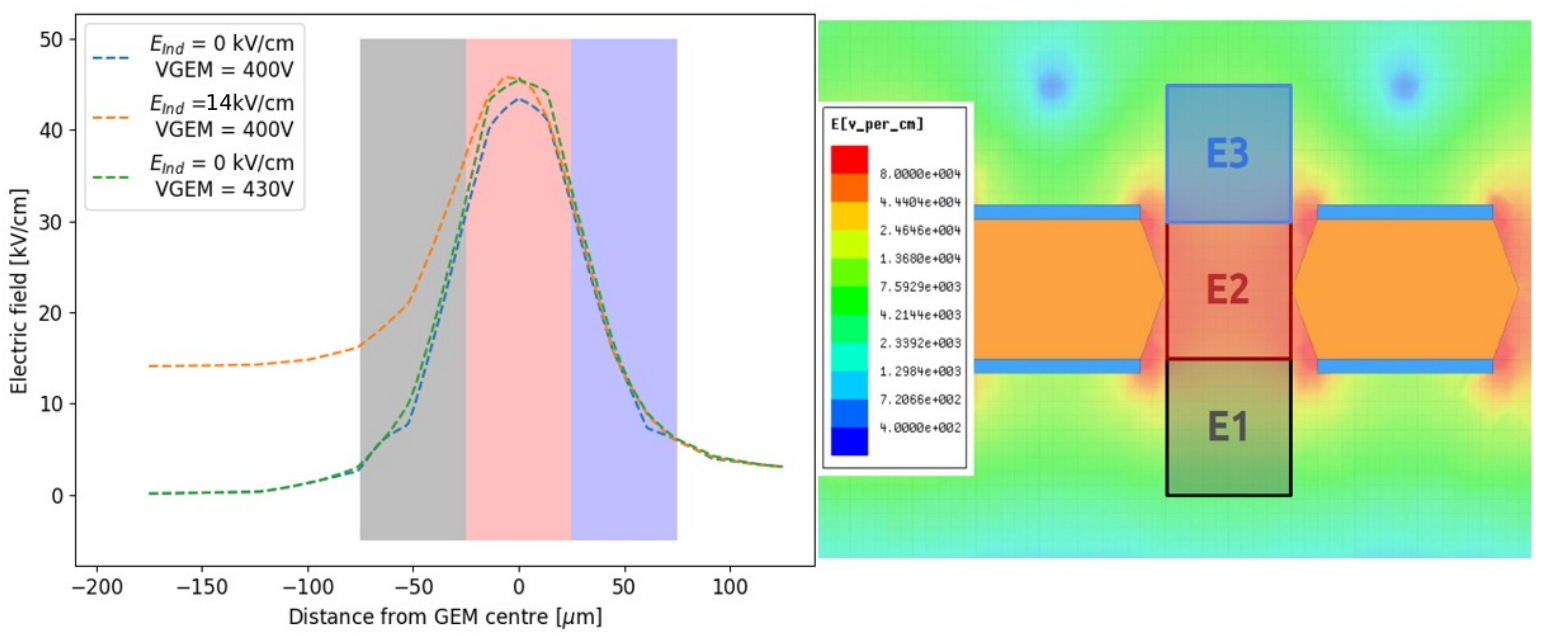}
	\caption{On the left, the profile of the electric field along the direction orthogonal to the GEM plane which passes through a $t$ GEM hole. The x-axis coordinate refers to the distance from the centre of the GEM hole, positive for above the GEM hole, negative for below, i.e. towards the induction gap. Three voltage configurations are depicted and described by the legend. Three regions are  highlighted in grey (E1), red (E2) and blue (E3), which are described in the text. On the right, a detail of the schematics of the $t$ GEM simulation with superimposed the same three regions E1, E2 and E3 described in the text.}
	\label{fig:profthin}
\end{figure*}
A generalised simulation of a generic TPC, identical to the internal structure of our prototype MANGO (see Fig. \ref{fig:lemonsketch} for a schematic sketch and Fig. \ref{fig:mangosketch} for the detector drawings), was initially performed with a coarse granularity. A copper 10$\times$10 cm$^2$ cathode encases with the amplification stage the drift volume with a 0.8 cm drift gap. The amplification stage comprises three GEMs, numbered 1 to 3 from the closest to the drift region to the outermost. At a distance of 3 mm away from GEM3, a metallic electrode is placed in order to provide the induction field. This simulation confirmed the existence of a region few centimetres away from the borders where both the drift and the induction fields are uniform, and that the electric fields close to the holes of GEM3 are independent from the voltage configuration of GEM1 and GEM2. Therefore, to study in detail the influence of the induction field in the nearby of the outer most GEM holes, only a single 10 $\times$ 10 cm$^2$ GEM foil including all its holes with proper dimensions and conic shape, together with the induction electrode is simulated in 2D (to minimise CPU time and since the geometry can be assumed to possess a cylindrical symmetry) and discussed in the following. The granularity and accuracy of the simulation were increased up to a point where the electric field values reached an asymptote, not to have the result influenced by numerical errors.
Two types of GEMs are simulated: 
\begin{itemize}
    \item \emph{t}: a thin 50 $\mu$m GEM with 70 $\mu$m radius holes and 140 $\mu$m pitch
    \item \emph{T}: a thicker 125 $\mu$m GEM with 175 $\mu$m radius holes and 350 $\mu$m pitch
\end{itemize}
The voltages applied across these two types of GEM mirrors typical values these objects are operated at in the CYGNO context (see Tab. \ref{tab:app} in Sec. \ref{sec:MANGO}).\\
Close to the GEM hole structure, the field exhibits non-uniform patterns both above and below it. The spatial scale of these irregularities covers a region of roughly 40 $\mu$m (100 $\mu$m) above and below a $t$ ($T$) GEM hole, as shown in Fig. \ref{fig:scheme} on the left (right). From the field vectors evaluation, in the example of the $t$ GEM, displayed in Fig. \ref{fig:flines}, it is possible to notice that the presence of the induction field straightens the field vectors below the GEM. Fig. \ref{fig:flines} shows on the left the field vectors in case no induction field is applied, whilst on the right a small 1 kV/cm induction field is present. The straightening of the field vectors is clearly visible making the electric field structure below each GEM more ordered.\\

\begin{figure*}[!t] 
	\centering
    \vspace{-0.3cm}
	\includegraphics[width=0.49\linewidth]{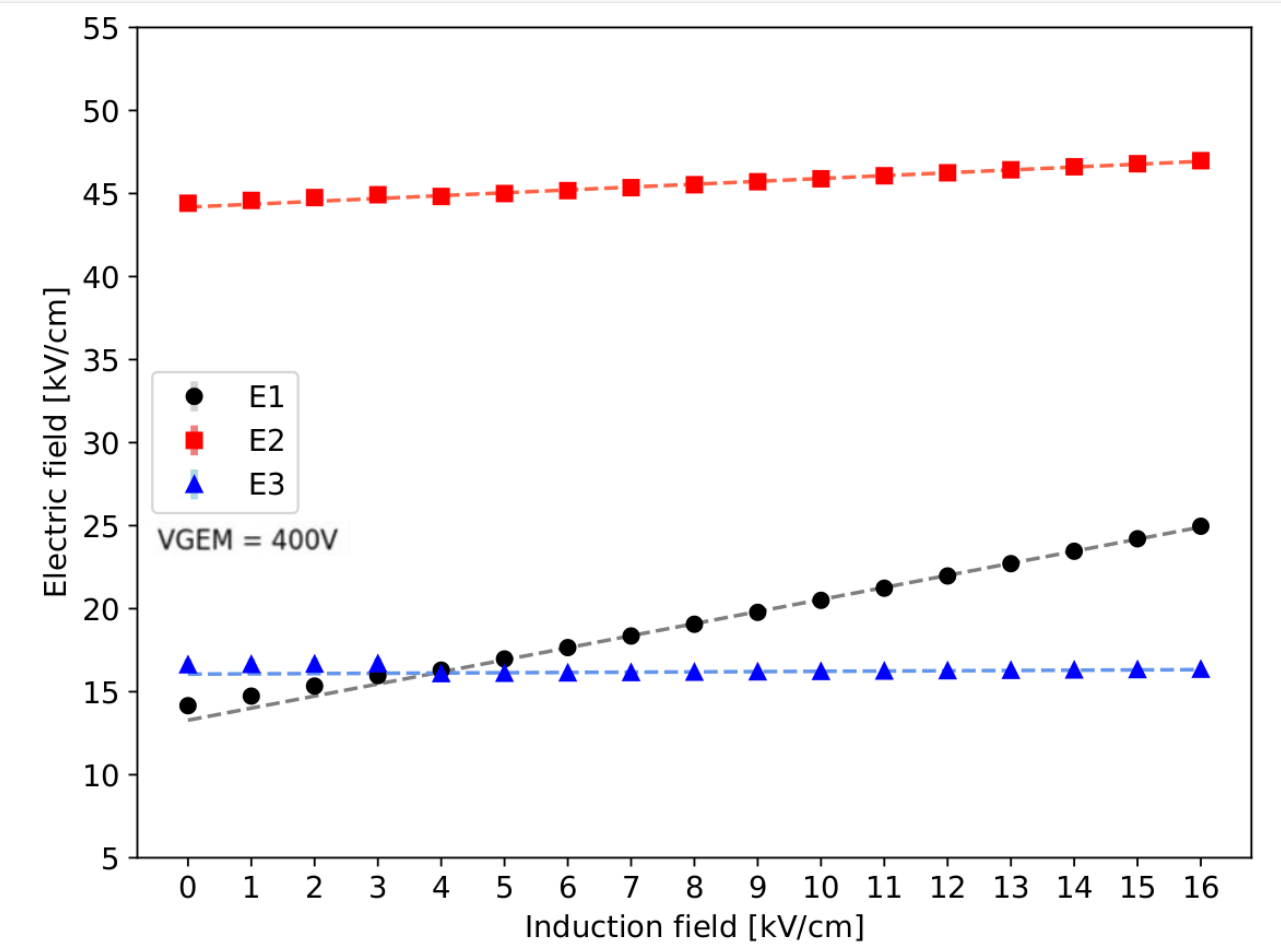}
	\includegraphics[width=0.49\linewidth]{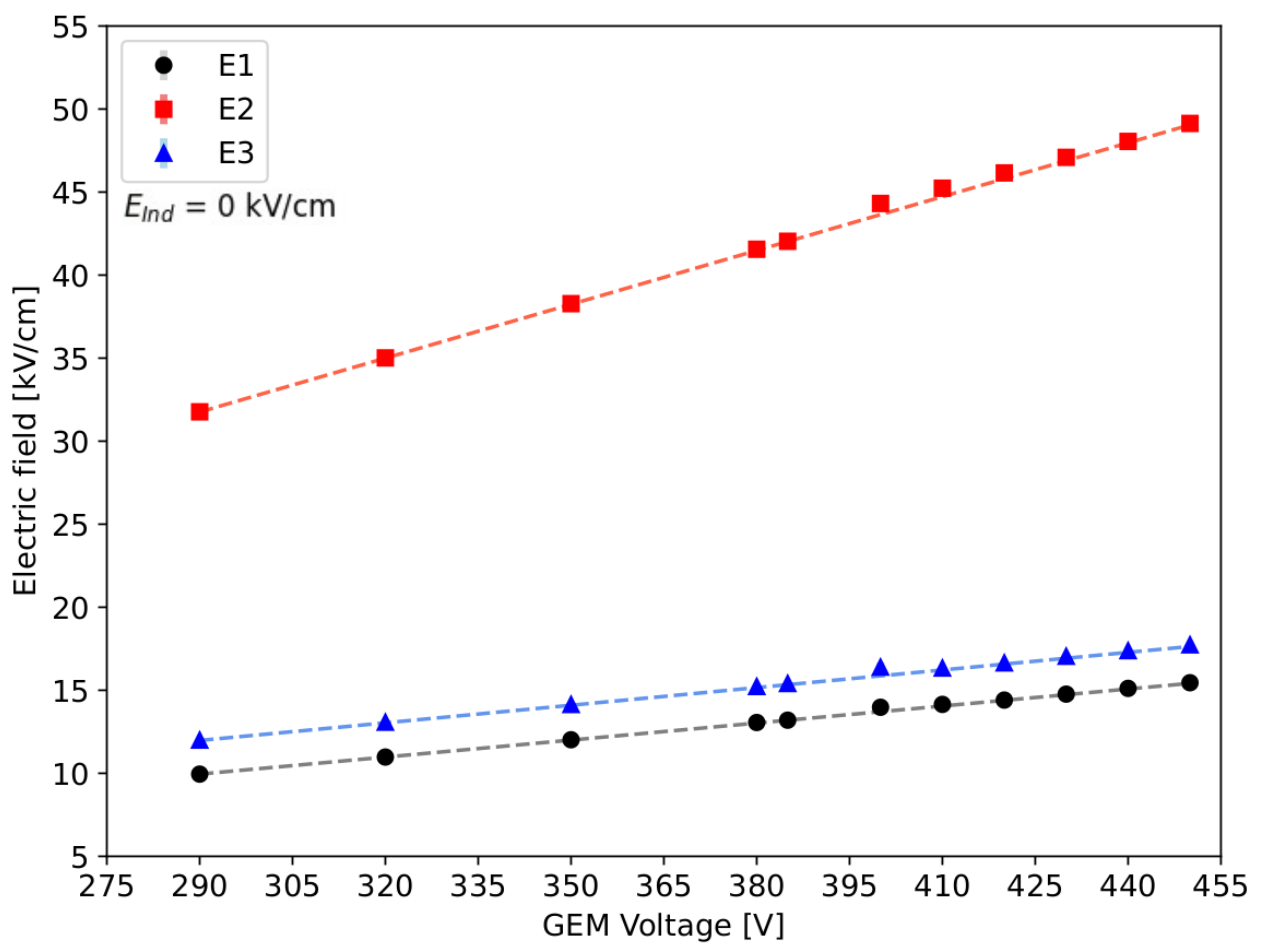}
	\caption{The simulated electric field in the three regions next to the GEM hole are displayed as a function of the induction field $E_{ind}$ on the left and as a function of V$_{GEM}$ on the right for a $t$ GEM geometry.}
	\label{fig:tttsim}
\end{figure*}
\begin{table*}[!t]
	\centering
	\begin{adjustbox}{max width=1.01\textwidth}
		\begin{tabular}{|c|c|c|c|c|}
			\hline
			Fit parameter & $A_{t,E}$ [kV/cm] & $B_{t,E}$ & $A_{t,V}$ [kV/cm] & $B_{t,V}$ [kV/V cm] \\
			\hline \hline
			E1 region     & 13.28 $\pm$ 0.03 & 0.727 $\pm$  0.003 & 0.043 $\pm$ 0.002 & 0.034 $\pm$ 0.001\\
			E2 region     & 44.18 $\pm$ 0.03 & 0.172 $\pm$  0.003 & 0.042 $\pm$ 0.006 & 0.108 $\pm$ 0.001\\
			E3 region     & 16.25 $\pm$ 0.06 & 0.004 $\pm$  0.005 & 1.72 $\pm$ 0.01 & 0.035 $\pm$ 0.001\\
			\hline
		\end{tabular}
	\end{adjustbox}
	\caption{Result of linear fits with the functions $A_{t,E}+ B_{t,E}E_{ind}$ and $A_{t,V}+B_{t,V}V_{GEM}$ to the electric fields in E1, E2 and E3 regions as simulated with Maxwell for a $t$ GEM. }
	\label{tab:tttsim}
\end{table*}
The profile of the electric field in the direction orthogonal to the GEM plane which passes through a $t$ GEM hole is shown in Fig. \ref{fig:profthin} on the left panel. The x-axis coordinate refers to the distance from the centre of the GEM hole, positive for above the GEM hole, negative for below, i.e. towards the induction gap. Three voltage configurations are shown: in blue 400 V across the GEM and no induction field is present, in green the voltage across the GEM is increased by 30 V, and finally in orange the voltage across the GEM is 400 but 14 kV/cm of induction field are present.
When no induction field is applied, the peak of the field is found at the zero coordinate, exactly at the centre of the GEM. The field symmetrically drops as the distance increases. Enlarging the voltage across the GEM affects the maximum field reached inside the GEM, but leaves the shape of the field profile untouched. Instead, when a strong induction field is added, not only the peak of the field inside the GEM hole increases, but the structure of the profile towards the induction gap changes. In particular, the decrease of the field has a milder slope, resulting in a stronger field close to the centre and edge of the GEM hole. In order to quantify this effect as a function of the voltage across the GEM and the induction field, the average value of the electric field is calculated in three different regions, highlighted in gray, red and blue in the left plot of Fig. \ref{fig:profthin}. The region E2 is a square of 50 $\times$ 50 $\mu$m$^2$ centred in the centre of the GEM foil, to characterise the field inside the GEM hole. The regions E3 and E1 are taken adjacent to E2, with the same dimensions, respectively above and below E2. Fig. \ref{fig:profthin} shows on the right panel the three regions in the Maxwell schematics.
The electric field in the three regions is simulated as a function of the induction field with constant 400 V applied across the GEM and as a function of V$_{GEM}$ with no induction field. The results are displayed in Fig. \ref{fig:tttsim} respectively on the left and right panel. For each value of $E_{ind}$ or V$_{GEM}$, the electric field simulated using Maxwell is averaged inside each box and the obtained mean is further averaged over 20 adjacent holes. Linear fits are performed on the sets of data and are summarised in Tab. \ref{tab:tttsim} for the induction field dependence, as $A_{t,E}+ B_{t,E}E_{ind}$, and for the V$_{GEM}$ one, as $A_{t,V}+B_{t,V}V_{GEM}$, where $A_{t,E},B_{t,E},A_{t,V},B_{t,V}$ are the fitting coefficients.
The results remark that increasing the voltage across the GEM modifies the field in all the three regions, symmetrically in E1 and E3, and with larger intensity in E2, as expected. Instead, the addition of the induction field augments the field in E1 more strongly than in E2, while E3 is left unaffected. This allows to conclude that it is impossible to increase the scintillation output by a modification of the GEM transparency due to the addition of an induction field underneath.
The electric field inside the GEM hole (E2 region) increases linearly with both the induction field and $V_{GEM}$, with a much stronger dependence on $V_{GEM}$. Nevertheless, in the E1 region, the fields reach values above 20 kV/cm, high enough to attain further amplification. Taking two configurations exemplified in Fig. \ref{fig:profthin}, namely where $V_{GEM}=$ 400 V and $E_{ind}=$ 14.0 kV/cm and where $V_{GEM}=$ 430 V and $E_{ind}=$ 0 kV/cm, a numerical comparison can be performed on the average intensity of the fields displayed in Fig. \ref{fig:tttsim}. It can be observed that the increase in the E1 field with respect to nominal operating conditions ($V_{GEM}=$ 400 V and $E_{ind}=$ 0 kV/cm) is a factor ten larger in the first case compared with the second one. Conversely, the increase in E2 is only a factor of 2 larger when $V_{GEM}$ is increased, with respect to a raise in $E_{ind}$. It also has to be noted that while this is an average value of the field, Fig. \ref{fig:profthin} shows that the closer one gets to the GEM, the larger the field intensity. Both Fig. \ref{fig:profthin} and Fig. \ref{fig:tttsim} also demonstrate that the high intensity of the electric field in region E1 is peculiar to the introduction of the induction field, thus it is not present in the regular operation of a GEM.\\

\begin{figure}[!t] 
	\centering
    \vspace{-0.5cm}
	\includegraphics[width=1.1\linewidth]{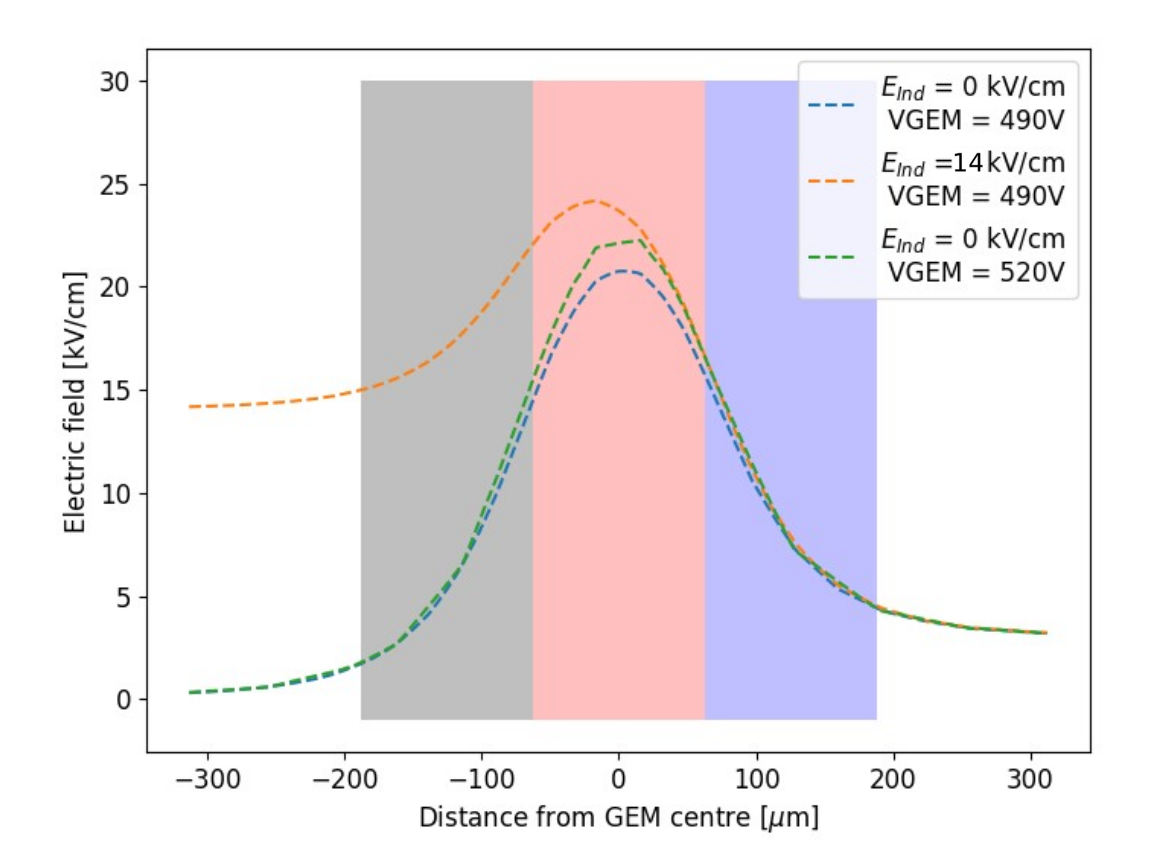}
	\caption{Profile of the electric field along the direction orthogonal to the GEM plane which passes through a $T$ GEM hole. The x-axis coordinate refers to the distance from the centre of the GEM hole, positive for above the GEM hole, negative for below, i.e. towards the induction gap. Three voltage configurations are shown as described by the legend. Three regions are  highlighted in grey (E1), red (E2) and blue (E3) which are described in the text.}
	\label{fig:profthick}
\end{figure}
\begin{figure*}[!t] 
	\centering
	\includegraphics[width=0.49\linewidth]{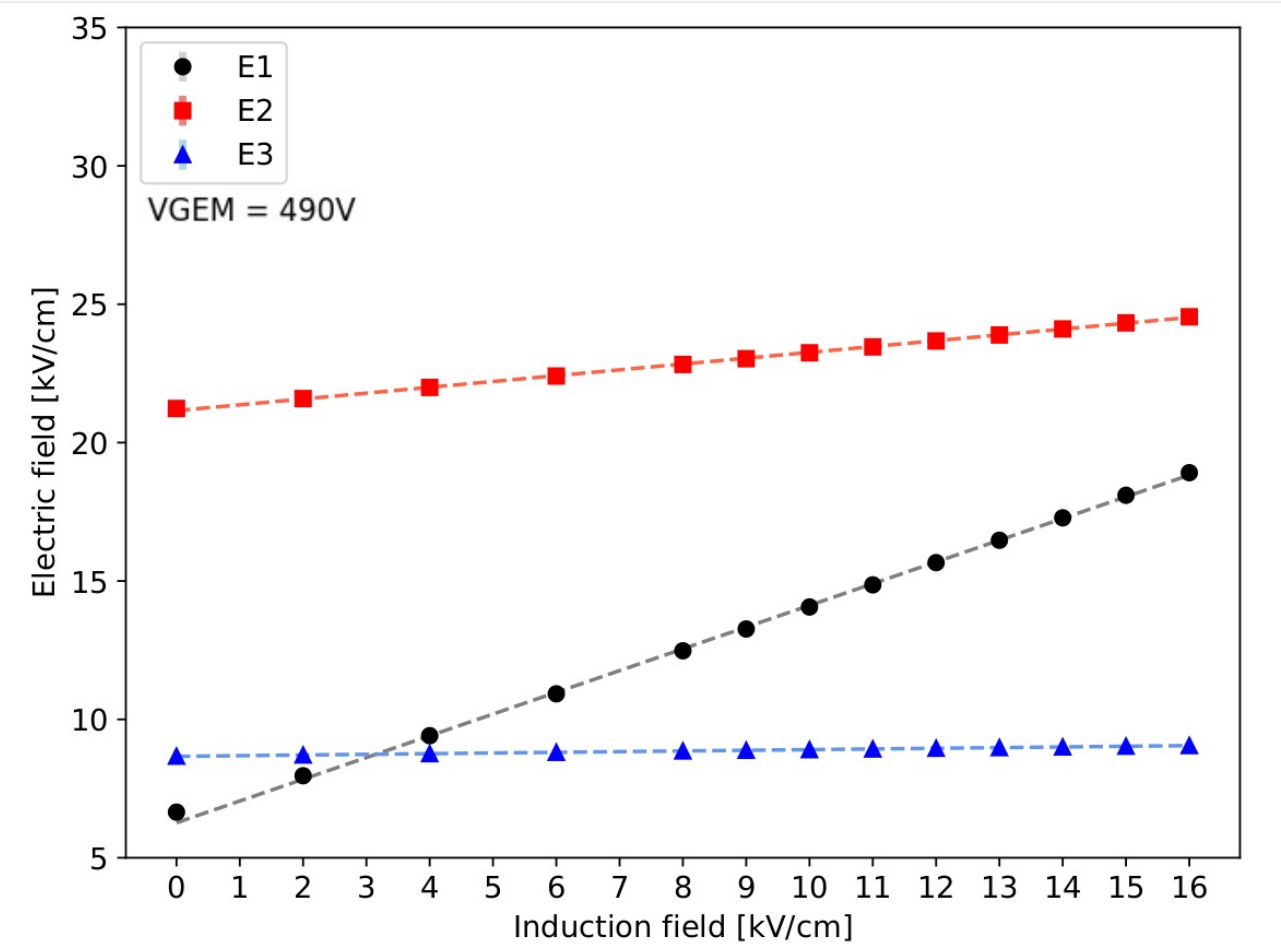}
	\includegraphics[width=0.49\linewidth]{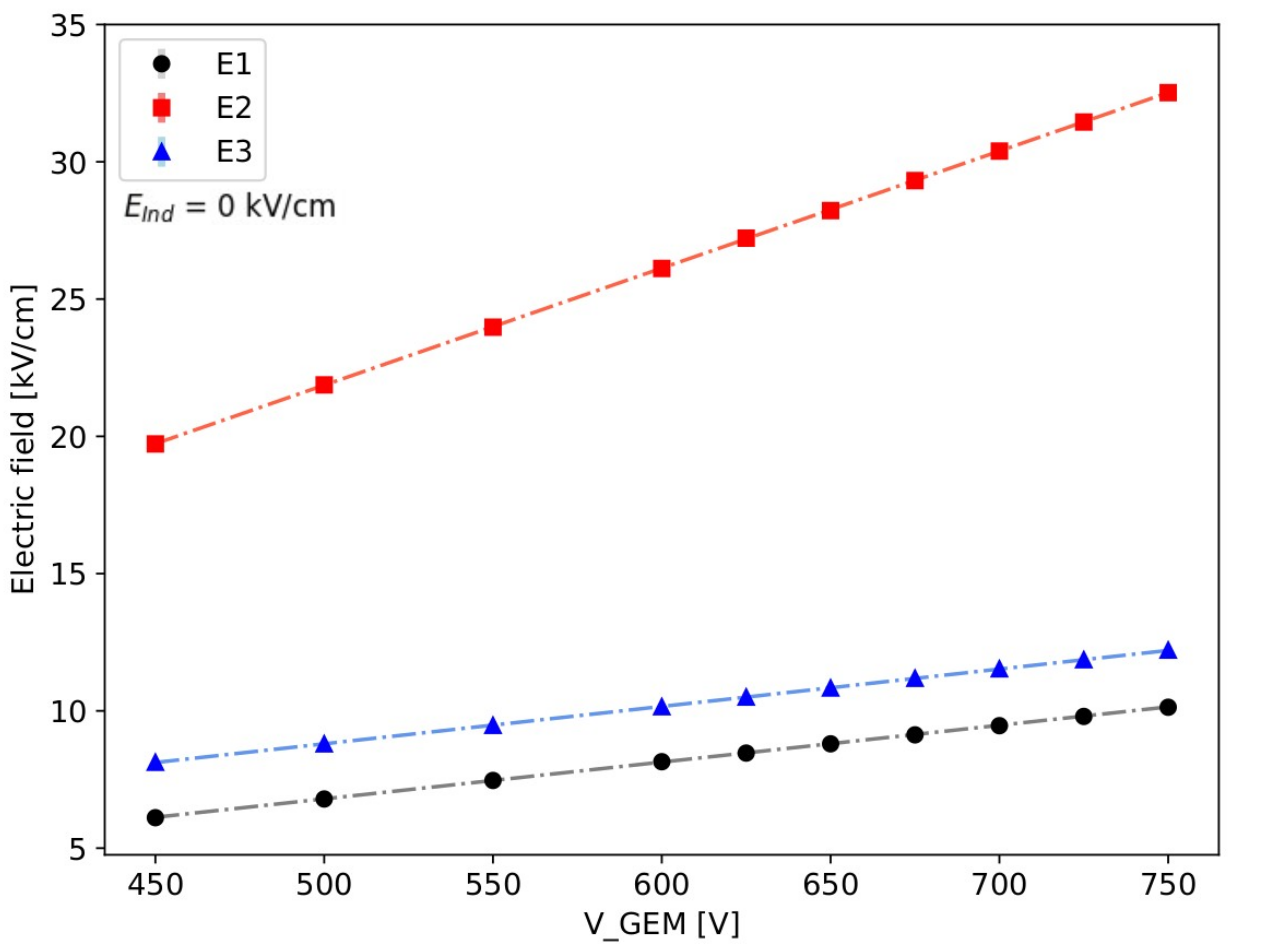}
	\caption{The simulated electric field in the three regions next to the GEM hole are displayed as a function of the induction field $E_{ind}$ on the left and as a function of V$_{GEM}$ on the right for a $T$ GEM geometry.}
	\label{fig:TTsim}
\end{figure*}
\begin{table*}[!t]
	\centering
	\begin{adjustbox}{max width=1.01\textwidth}
		\begin{tabular}{|c|c|c|c|c|}
			\hline
			Fit parameter & $A_{T,E}$ [kV/cm] & $B_{T,E}$ & $A_{T,V}$ [kV/cm] & $B_{T,V}$ [kV/V cm] \\
			\hline \hline
			E1 region     & 6.27 $\pm$ 0.02 & 0.785 $\pm$  0.002 & 0.09 $\pm$ 0.06 & 0.0134 $\pm$ 0.0005\\
			E2 region     & 21.16 $\pm$ 0.03 & 0.210 $\pm$  0.001 & 0.54 $\pm$ 0.06 & 0.042 $\pm$ 0.001\\
			E3 region     & 8.66 $\pm$ 0.02 & 0.002 $\pm$  0.006 & 2.00 $\pm$ 0.03 & 0.0136 $\pm$ 0.0005\\
			\hline
		\end{tabular}
	\end{adjustbox}
	\caption{Result of linear fits with the functions $A_{T,E}+ B_{T,E}E_{ind}$ and $A_{T,V}+B_{T,V}V_{GEM}$ to the electric fields in E1, E2 and E3 regions as simulated with Maxwell for a $T$ GEM. }
	\label{tab:TTsim}
\end{table*}
The same type of simulation and analysis is performed on a $T$ GEM, with a reference voltage across it taken as 490 V (see Tab. \ref{tab:app}, Sec. \ref{sec:MANGO}). Fig. \ref{fig:profthick} shows the profile of the electric field along the direction orthogonal to the GEM plane which passes through a $T$ GEM hole. The modification of the electric field structure, with respect to the reference 490 V applied across the GEM, are obtained by raising the voltage of 30 V or by introducing 14 kV/cm in the induction gap. Akin to the $t$ GEM, when the induction field is added, not only the maximum field reached inside the GEM increases, but the variation of the field intensity with the distance from the GEM has a harder slope with respect to when only the GEM voltage is raised.  Due to the geometry when compared to the $t$ GEM case, the distortion induced by the induction field acts on a wider area below the GEM hole and has a larger relative impact on the field intensity. To quantify the influence of the variation of the electric field as a function of the voltage across the GEM and the induction field, three regions are defined in the 2D space of the simulation the average value of the electric field is calculated from, exactly as for the $t$ GEM. In order to adapt to a larger $T$ GEM the regions are selected with an area of 125 $\times$ 125 $\mu$m$^2$.\\
The results of the average field in the E1, E2, and E3 regions as a function of the induction field and V$_{GEM}$ are displayed in Fig. \ref{fig:TTsim} respectively on the left and right panel. Linear fits are performed on the sets of data and are summarised in Tab. \ref{tab:TTsim} for the induction field dependence, as $A_{T,E}+ B_{T,E}E_{ind}$, and for the V$_{GEM}$ one, as $A_{T,V}+B_{T,V}V_{GEM}$ - where $A_{T,E},B_{T,E}$ $,A_{T,V},B_{T,V}$ are the fitting coefficients.\\
The results for the $T$ GEM are coherent to the one attained for the $t$ GEM. The electric fields are generally lower in the case of the $T$ GEM, as expected from the dimension of the holes, and the values obtained in the region E2 of $\sim$ 20 kV/cm confirm that these fields allow amplification processes, as it will be seen in  Tab. \ref{tab:app} (Sec. \ref{sec:MANGO}) and Fig. \ref{fig:gain} (Sec. \ref{subsec:mango_light}). Nonetheless, it can be noted that the slope of the increase of the field is larger for the $T$ GEM than the $t$ GEM one.\\

The results of these simulations show that an increase in light production is possible due to a increase of the field inside the GEM holes. More interestingly though, strong induction fields can generate a region towards the bottom of the GEM hole and right below it where amplification and photon creation is possible. Given the physical dimension of this region and field intensity, it is possible that abundant excess of light is produced. In particular, given the typical lower electric fields inside the holes and because of the larger size and electric field influence region, a $T$ GEM is expected to undergo stronger light enhancement with respect to a $t$ GEM. As a consequence, an experimental detailed investigation of this process is worth to be performed.

\section{LEMOn experimental setup}
\label{sec:lemon}
\label{subsec:lemon_det}
\begin{figure}[!t] 
	\centering
	\includegraphics[width=0.9\linewidth]{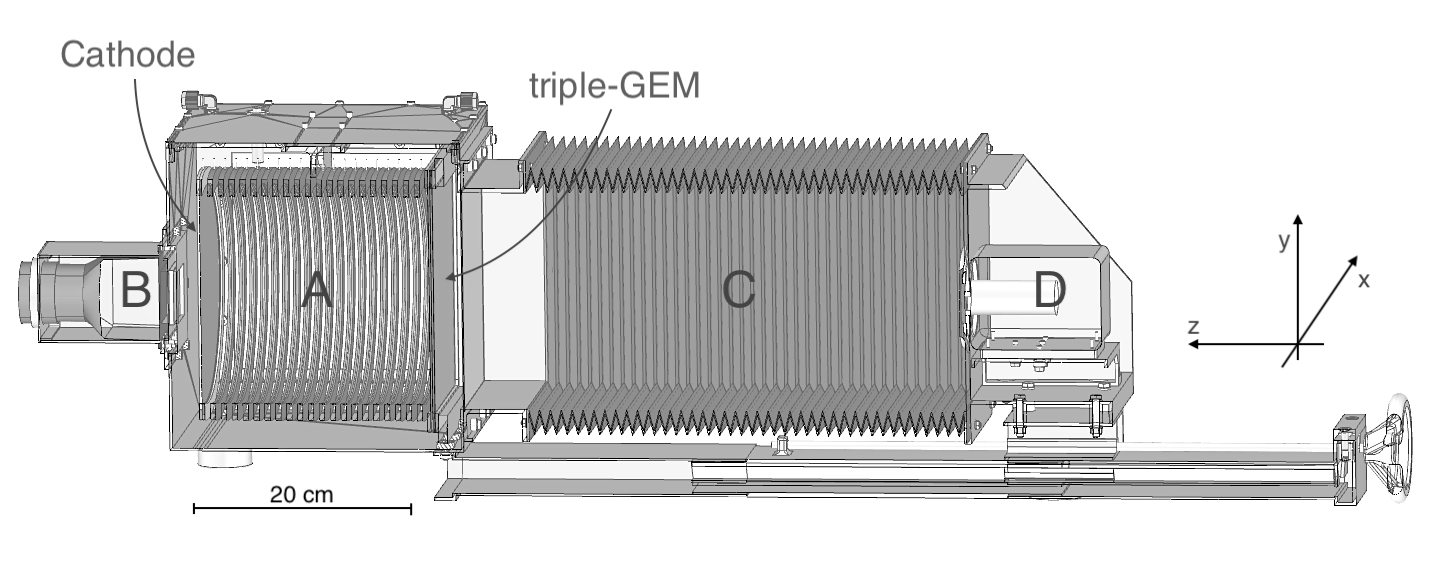}
 	\includegraphics[width=0.4\textwidth]{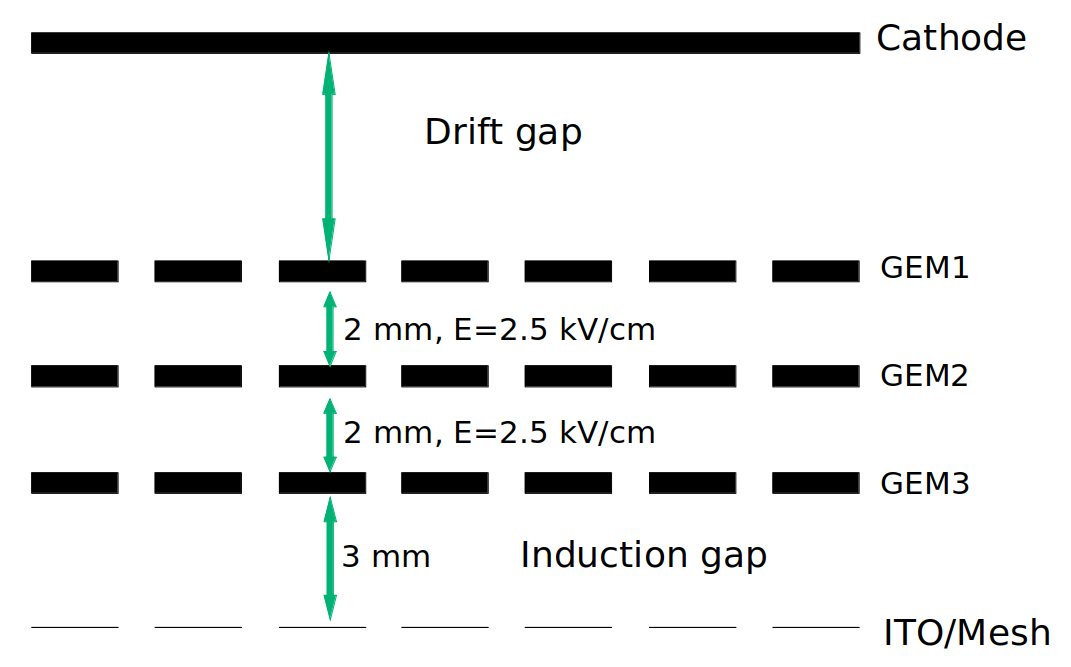}
	\caption{On the top, the LEMOn prototype \cite{bib:Antochi_2021}. The~elliptical sensitive volume (\textbf{A}), the~fast photo-multiplier (\textbf{B}), the~optical bellow (\textbf{C}) and the sCMOS-based camera (\textbf{D}) are~indicated. On the bottom, a sketch of the internal structure of the TPC of the CYGNO prototypes employed in this study where the addition of the ITO or a mesh below the last amplification GEM plane can be appreciated.}
	\label{fig:lemonsketch}
\end{figure}

In order to test the results of the simulation presented in the previous Section, to further validate the findings we presented in \cite{bib:EL_cygno} as well as to extend them to stronger applied electric fields, we employed a larger detector, the Long Elliptical MOdule (LEMOn).\\
A sketch of the LEMOn detector is shown in Fig. \ref{fig:lemonsketch} on the top panel. A 7 litres active volume TPC with a 20 cm drift length and a 24 $\times$ 20 cm$^2$ readout area is enclosed in a gas-tight acrylic vessel and operated in continuous gas flux mode. An ellipsoidal field cage comprised of silver wires held by 3D printed plastic supports with 1 cm pitch guarantees drift field uniformity in the 20 cm drift gap. The cathode is manufactured from an ATLAS MicroMegas mesh \cite{bib:Micromegas} with 30 $\mu$m diameter metallic wires with a pitch of 70 $\mu$m. To generate electron avalanche at the amplification stage a geometry similar to the one described in Sec. \ref{sec:maxwell} is employed with three 24 $\times$ 20 cm$^2$ GEMs, 50 $\mu$m thick, 70 $\mu$m hole diameter and 140 $\mu$m pitch, spaced each other 2 mm.
The GEMs are numbered from 1 to 3 with GEM1 being the closest to the drift region. At a distance $\Delta z$ of 3 mm below GEM3, an ITO glass electrode with a transparency of 90$\%$ is placed in order to establish an additional electric field in the region below the last GEM (GEM3) to enhance the light yield of the detector. This region between GEM3 and the ITO glass is the induction region. A sketch of the internal structure of the TPC is displayed on the bottom panel of Fig. \ref{fig:lemonsketch}. LEMOn cathode and field cage base are powered by a CAEN N1570\footnote{\url{https://www.caen.it/products/n1570/}} HV supply, while the GEMs are biased by a CAEN A1526\footnote{\url{https://www.caen.it/products/a1526/}} power supply with 6 independent HV channels up to 15 kV with a current sensitivity of 10 nA. The latter  provides a very precise tool of measurement of the charge on each of the GEM electrodes for currents above tens of nA. The ITO electrode is biased by a CAEN DT1470ET\footnote{\url{https://www.caen.it/products/dt1470et/}} power supply with a high sensitivity current-meter ($\sim$ 5 nA) able to precisely measure the continuous current signals.

The LEMOn detector is optically coupled to a Hamamatsu sCMOS camera (C14440-20UP ORCA-Fusion) through a TEDLAR transparent window and an adjustable plastic bellow. The camera is equipped with a Schneider Xenon lens with 25.6 mm focal length and 0.95 aperture. The Orca Fusion is positioned at $(50.6 \pm 0.1)$ cm distance from GEM3 and reads out an area of 25.6  $\times$ 25.6 cm$^2$. Therefore, each of the 2304 $\times$ 2304 pixels of the sCMOS sensor images an effective area of 111  $\times$ 111 um$^2$. A more detailed description of LEMOn and its performances can be found in \cite{bib:Antochi_2021,bib:fe55}.

In this setup LEMOn is operated with a 0.5 kV/cm drift field, 2.5 kV/cm transfer fields between GEMs and 400 V applied across each of them, with a He:CF$_4$ 60/40 gas mixture at 1000 mbar, as it is installed at Laboratori Nazionali di Frascati (LNF). A $\sim$ 115 MBq $^{55}$Fe source is used to induce 5.9 keV energy deposits inside the detector active gas volume and it consists in a small cylinder (1 cm in height and 0.2 cm diameter) made of iron where only one tip is radioactive. It is located right outside the gas volume at 5 cm distance from GEM1 facing a \myl window not to affect the electric field of the detector and to allow the X-rays to reach the active volume of the TPC. The large source activity provides a detectable current signal on each of the 3 GEM electrodes and the ITO glass, given the current sensitivity of the supply configuration. Conversely, the source intensity does not allow to identify each $^{55}$Fe cluster separately because of the large pileup. For this reason, and since the current on the electrodes represents an integrated information of all the $^{55}$Fe clusters produced in the gas and amplified by the GEMs, a 1 s exposure time on self-trigger pictures is used for the sCMOS camera data acquisition in LEMOn to perform a consistent light measurement. The light yield from the sCMOS camera and the charge measured on each of the seven LEMOn amplification electrodes (one for the ITO and two for each GEM, the upper (U) and the bottom (D) ones) are studied by varying the induction field E$_{ind}$ from 0 to 17 kV/cm for a constant V across each GEM and the results are reported in the following.

\subsection{sCMOS images analysis}\label{sec:lemon_light}
\begin{figure}[t!] 
	\centering
	\includegraphics[width=0.7\linewidth]{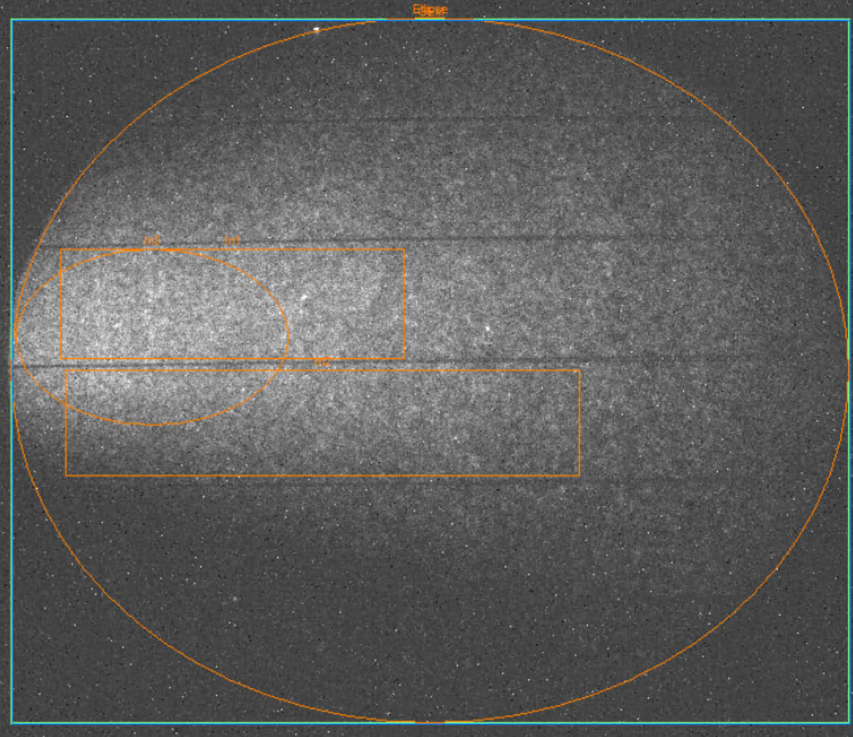}
	\caption{Example of 1 s exposure picture taken with the sCMOS camera with superimposed three regions the total light was evaluated from, as described in the text in Sec. \ref{sec:lemon_light}.}
	\label{fig:longexplemon}
\end{figure}
An example of a 1 s exposure sCMOS image acquired by LEMOn exposed to the $^{55}$Fe source is shown in Fig. \ref{fig:longexplemon} with highlighted four different regions (two elliptical and two rectangular) in orange. The light yield in the LEMOn data is evaluated by calculating the number of counts seen by all the sCMOS pixels in each region, after having subtracted the noise pixel by pixel exploiting images acquired in absence of source and with the GEM turned off. This number is normalised to the measurement with null induction field and the relative increase is averaged among the four regions.
Fig. \ref{fig:light_charge_lemon} shows the comparison of the relative increase of the light output and the charge measured on the ITO glass (see Sec. \ref{subsec:lemon_charge} for details on how the charge is evaluated) as a function of the induction field in He:CF$_4$ 60/40 at 1000 mbar, explicitly demonstrating the different rate of increase of the two quantities. The light enhancement measured with LEMOn and shown in Fig. \ref{fig:light_charge_lemon} is consistent with the results reported in \cite{bib:EL_cygno} when one considers the errata corrige to the induction gap dimension claimed in that paper (actual: 2.5 mm, instead of the claimed 3 mm).

\subsection{GEMs electrodes current analysis}
\label{subsec:lemon_charge}
During the amplification processes a large amount of electrons and ions are generated close to the bottom of the holes of each GEM and drifted away in opposite directions by the electric fields applied. Considering the Shockley-Ramo theorem \cite{bib:Ramo,bib:Shockley}, the instantaneous current induced on the GEM electrodes depends on the amount of charge in motion (both ions and electrons), on their velocity and on a function of the electric field along the charge path from its generation to the point of collection. The significant difference in ion and electron drift velocity leads to an average charge collection time of the order of $\mu$s for the former and few ns for the latter.
The signal is induced as soon as a charge gets in motion and lasts until it is collected by an electrode. It is important to notice how a current signal is induced also on electrodes not collecting any charge due to the electron and ion motion. In this case, the signal is bipolar and its overall integral sums to zero. As a consequence, an infinite integration of the current signal allows to correctly measure a signal dependent solely on the actual charge collected by an electrode.
\begin{figure}[!t] 
	\centering
	\includegraphics[width=1.1\linewidth]{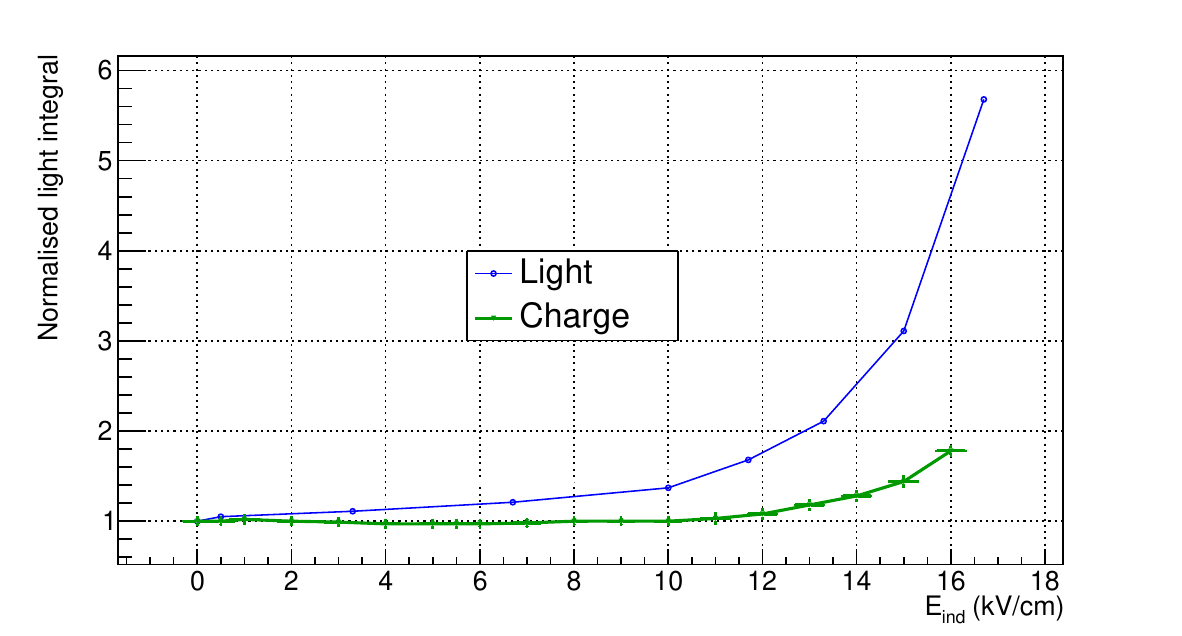}
	\caption{Comparison of the relative increase of light and charge integral for LEMOn in He:CF$_4$ 60/40 at 1000 mbar.}
	\label{fig:light_charge_lemon}
\end{figure}

Since no charge needs to be collected with the optical readout approach, all CYGNO prototypes are operated with the GEM3 bottom electrode at ground. In this configuration, the absence of induction field E$_{ind}$ causes the field lines to weakly and disorderly close on the bottom electrode of the GEM. Immediately after the multiplication inside the holes of the last GEM, an electric signal is induced on its electrodes. The upper electrode is responsible for the collection of the majority of the ions coming from the last step of multiplication, while the bottom one of the electrons. When the induction field E$_{ind}$ is turned on, the field lines are straightened and begin to close on the ITO glass rather than on the bottom GEM3 electrode. In this configuration, the ions are expected to be mostly collected on the top of GEM3, while the electrons will be shared between the bottom of GEM3 and the ITO glass. When the electric field inside the induction gap is large enough to generate charge amplification, these additional electrons are collected on the ITO, while the newly generated ions are shared between the other electrodes, with the large majority of them being collected by the top and bottom ones of GEM3.
\begin{figure*}[!t] 
	\centering
	\includegraphics[width=0.8\linewidth]{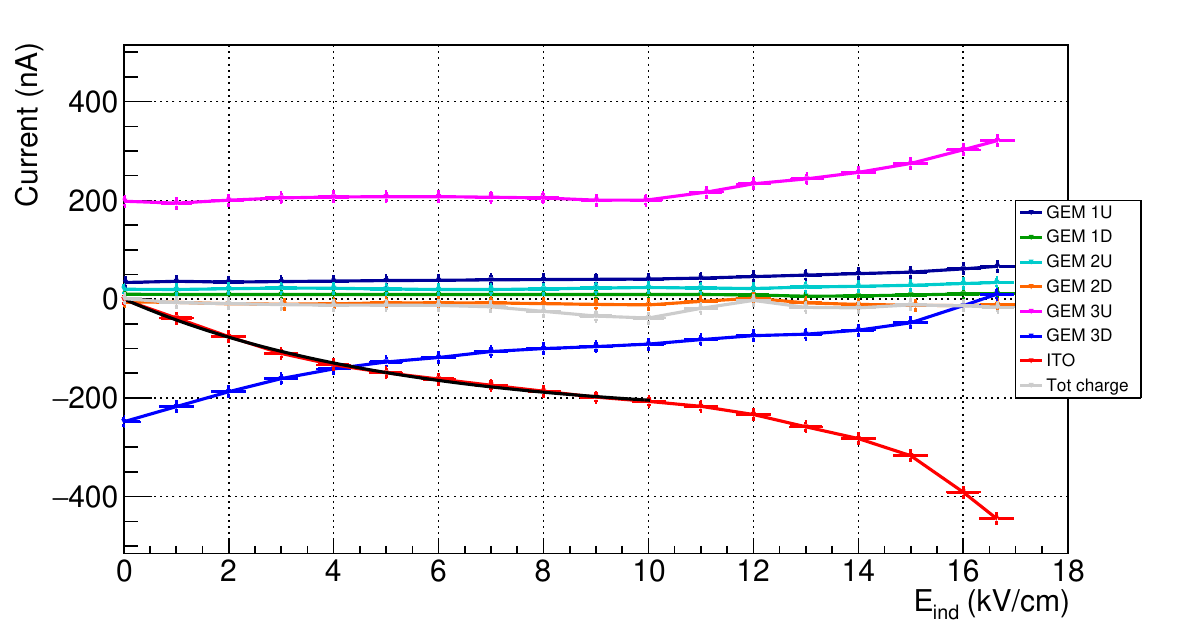}
	\caption{Currents measured in LEMOn as a function of the induction field E$_{ind}$ for all the six electrodes of the amplification stage plus the ITO glass, where \emph{U} and \emph{D} represent respectively the upper and the bottom electrode of each GEM, and the total charge (in gray) is the sum of all the components. The black line represents the exponential fit to the ITO curve described in Equation \ref{eq:fit}.}
	\label{fig:Lemoncharge}
\end{figure*}
Given the LEMOn setup, the currents measured in this context can be considered as equivalent to an infinite integration, also considering the large $^{55}$Fe source activity employed. Fig. \ref{fig:Lemoncharge} shows the continuous current measured in LEMOn as a function of the induction field for all the six electrodes of the amplification stage plus the ITO glass, and the total charge (in gray) is the sum of all the components.

The sum of the charge of all electrodes is always consistent with zero and mostly flat, as expected from general arguments by a well grounded electrical circuit. Similarly, the ITO glass and GEM 3D split between them (with proportions depending on the induction field) the current induced by the charges generated in the last amplification stage and moving in the induction gap. When no induction field is applied, the ITO sees a null current and all the 250 nA are collected at GEM 3D. The current measured on the upper electrode of the last GEM, GEM 3U, as a function of E$_{ind}$ displays a constant behaviour up to about 10 kV/cm, where an increase starts. This \emph{breaking} point behaviour at 10 kV/cm is shared with the ITO and GEM 3D measurements, which show an inflection point at the same value.
In order to evaluate the actual relative increment of measured charge with respect to null induction field, the charge sharing between GEM 3D and the ITO glass needs to be properly taken into account. To do this, the ITO current measured for induction fields between 0 and 10 kV/cm is fitted with the function (black line in Fig. \ref{fig:Lemoncharge}):
\begin{equation}
\label{eq:fit}
I_{ITO} = a+e^{b+cE_{ind}}.
\end{equation}
The following parameters are obtained: $a= (-240 \pm 20)$ nA, $b= (5.47 \pm 0.08)$ , $c= (-0.20 \pm 0.05)$ cm/kV. The resulting parameter $a$, which represents the asymptote of the exponential function, exhibits a good agreement with the -250 nA value measured on GEM 3D due to the collection of all the electrons generated in the last GEM when no induction field is applied. In order to properly evaluate the actual charge generated in the induction gap, the ITO data are normalised to the value at 10 kV/cm after having subtracted the fitted  $I_{ITO}$ function. These data are shown as a function of the induction field E$_{ind}$ in Fig. \ref{fig:light_charge_lemon} in comparison to the light output relative enhancement (illustrated in Sec. \ref{sec:lemon_light}), explicitly displaying the different derivative in increase of the two quantities. This increment above 10 kV/cm in the measured charge is attributed to the generation of a small additional amount of charge right below GEM3 holes, which, nonetheless, can not account for the entire increase of the light output. This is coherent with the discussion of Sec. \ref{sec:scintill} and \ref{sec:maxwell}. If the enhancement of light is produced at high E$_{ind}$ in the region underneath the GEM hole as suggested by the Maxwell simulations, the electric fields involved have lower intensities than the ones within the GEM holes. Thus, the light-producing process results favoured in terms of cross section with respect to charge production.

\section{MANGO Experimental setup} \label{sec:MANGO}
\begin{figure}[!t]
	\centering
	\includegraphics[width=0.95\linewidth]{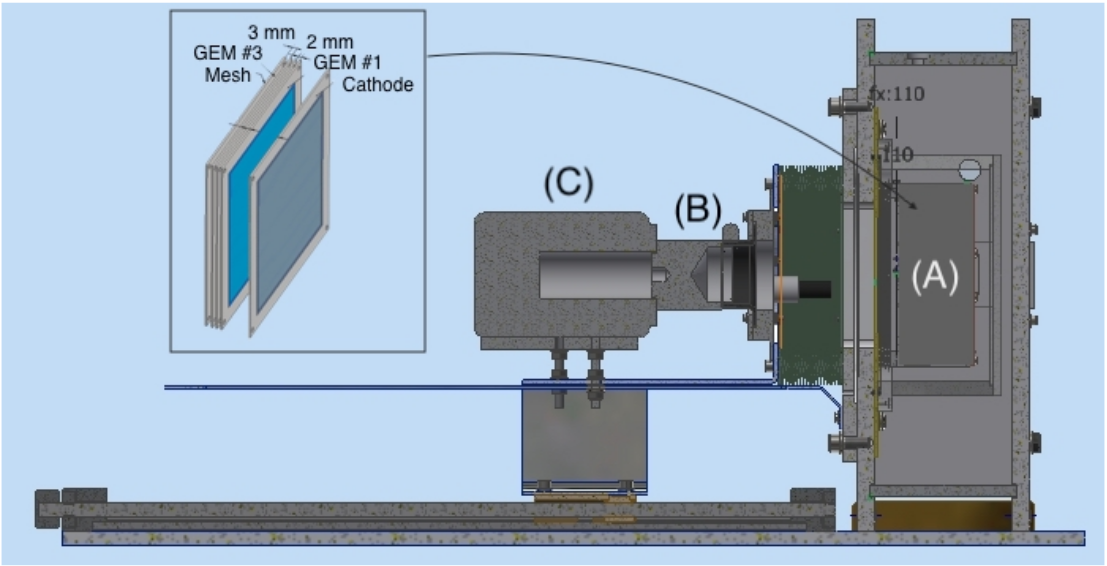}
	\caption{A simple representation of the MANGO setup with exemplified triple GEM amplification.}\label{fig:mangosketch}
\end{figure}
\begin{table*}[!t]
	\centering
	\begin{adjustbox}{max width=1.01\textwidth}
		\begin{tabular}{|c|c|c|c|c|c|c|c|}
			\hline
			
			\multirow{2}{*}{\large{Ampl stage}} & \multirow{2}{*}{\large{Tag}} & \multicolumn{2}{c|}{\large{Gas mixture (He:CF$_4$)}}  &\multicolumn{4}{c|}{\large{Study}}    \\ \cline{3-8} 
			& & 60/40 & 70/30 &  Gain & Energy Res &  Diffusion & E$_{ind}$ \\ \hline \hline
			Triple thin GEM & ttt & $\checkmark$  &  & $\checkmark$ & $\checkmark$ & $\checkmark$ & $\checkmark$ \\ \hline
			Double thick GEM & TT & $\checkmark$   & $\checkmark$ & $\checkmark$ & $\checkmark$ & $\checkmark$ & $\checkmark$ \\ \hline
			1 thick and 1 thin GEM & Tt   & $\checkmark$ & $\checkmark$ & $\checkmark$ & $\checkmark$ & $\checkmark$ & $\checkmark$ \\ \hline
		\end{tabular}
	\end{adjustbox}
 \caption{Table summarising the gas mixtures and GEMs configurations explored in this study.}
	\label{tab:datataking}
\end{table*}
\begin{table*}[!t]
	\centering
	\begin{adjustbox}{max width=1.01\textwidth}
	\begin{tabular}{|c|c|c|c|c|}
		\hline
		\large{Config} 	& $V_{GEM1}$ [V] & $V_{GEM2}$ [V] & $V_{GEM3}$ [V]  & $V_{GEM}$ for E$_{ind}$ studies [V]\\ \hline \hline
		ttt 60/40 & 400-435  & 400-435  & 400-435 & 400+400+400=1200\\ 
		TT 60/40 & 770-780  & 470-520 & n.a & 775+490=1265\\ 
		Tt 60/40 & 740-780 & 400-435 & n.a & 770+400=1170\\ \hline
		TT 70/30 & 700-715 & 500 & n.a & 700+490=1190\\ 
		Tt 70/30 & 660-720 & 350-395 & n.a & 700+385=1085\\ \hline
	\end{tabular}
\end{adjustbox}
 \caption{Table summarising the voltages applied to the various combinations of GEMs structure explored in this study. Each column shows the range of voltages employed for each GEM.}
	\label{tab:app}
\end{table*}
In order to expand and extend the results we presented in \cite{bib:EL_cygno} and validated in Sec. \ref{sec:lemon}, and to further test the results of the Maxwell simulations presented in Sec. \ref{sec:maxwell}, we employed a smaller detector, namely the Multipurpose  Apparatus for Negative ions studies with GEM Optically readout (MANGO), to be able to modify the GEMs thicknesses and stacking option, in addition to the gas mixture (since no 20 $\times$ 24 GEMs are available with thickness different from the standard 50 $\mu$m).

A sketch of the MANGO prototype is shown in Fig. \ref{fig:mangosketch} and described in more details in \cite{bib:EL_cygno}, while the internal TPC structure used for these measurement is similar to the one displayed in Fig. \ref{fig:lemonsketch} on the right. The amplification stage consists in a stack of multiple GEMs of 10 $\times$ 10 cm$^2$, spaced 2 mm, with a transfer field of 2.5 kV/cm in between. The number and type of GEMs used changed during the data taking to explore the performances of different thickness and multiple stacking options. At a distance $\Delta$z$ =3$ mm a metallic mesh from an ATLAS MicroMegas \cite{bib:Micromegas} (30 $\mu$m diameter metallic wires at 50 $\mu$m pitch, resulting in a transparency of $\sim 0.55 $) is placed in order to induce an electric field below the electrode of the last GEM. A different configuration was also tested, replacing the metallic mesh with an ITO glass, similar to the one employed in LEMOn, with a larger transparency (0.9). This test showed that the results do not depend on the structure employed to produce the additional electric field after the last GEM amplification. As in LEMOn and in the Maxwell simulations, we define the region between the last GEM amplification plane and the mesh as the \emph{induction} region, and the electric field applied inside it the \emph{induction} field (E$_{ind}$).

The drift gap measures 0.8 cm and the detector is operated with 1 kV/cm drift field, a configuration that guarantees a uniform electric field in the drift region without the need for a field cage. The TPC structure is enclosed in a 3D printed black plastic light-tight box that contains a gas-tight acrylic internal vessel. A thin window of highly transparent (> 0.9) Mylar\textregistered\xspace decouples the gas detector from the optical readout, which consists in a PMT (Hamamatsu H3164-10) and the same ORCA FusionsCMOS camera as in LEMOn detector (C in Fig. \ref{fig:mangosketch} left), placed at a distance of $(20.5 \pm 0.3)$ cm and focused on the last GEM ampilfication plane. The camera is equipped with the same lens as in the LEMOn prototype. Within this scheme, the camera images an area of 11.3 $\times$ 11.3 cm$^2$, resulting in an effective pixel size of 49 $\times$ 49  $\mu$m$^2$. Various combinations of He:CF$_4$ mixtures are used in this study, always keeping the helium content above 60\%. All of the mixtures used scintillate with peaks at 620 nm as described in Sec. \ref{sec:scintill}, where the quantum efficiency of the camera sensor reaches 0.8 and the Schneider lens transparency about 0.85.

\subsection{Datasets}
\label{subsubsec:mango_daq}
A $\sim$ 480 kBq $^{55}$Fe X-rays source is employed to generate 5.9 keV signals in the MANGO active gas volume. The relative low source activity allows the reconstruction of each single $^{55}$Fe signal in the sCMOS images (acquired with 0.5 s exposure), as discussed in Sec. \ref{subsec:mango_light} and shown in Fig. \ref{fig:waveformGEM}.

A systematic study of the performances of different He:CF$_4$ ratios in the gas mixtures, 60/40 and 70/30, and different GEM thicknesses and stacking options is performed. The same two types of GEM described in the simulation Section are employed: a thin 50 $\mu$m GEM with 70 $\mu$m radius holes and 140 $\mu$m pitch ($t$) and a thicker 125 $\mu$m GEM with 175 $\mu$m radius holes and 350 $\mu$m pitch ($T$). All the measurements are performed at the atmospheric pressure at Laboratori Nazionali del Gran Sasso (LNGS) - located at roughly 1000 m a.s.l. - which corresponds to (900 $\pm$ 7) mbar.

Tab. \ref{tab:datataking} shows a summary of the different gas mixtures and GEM configurations explored with MANGO, and Tab. \ref{tab:app} the voltages applied to the various combinations of GEM structures.

\section{sCMOS images analysis}
\label{subsec:mango_light}
\begin{figure}[t] 
	\centering
 \vspace{-0.1cm}
	\includegraphics[width=0.65\linewidth]{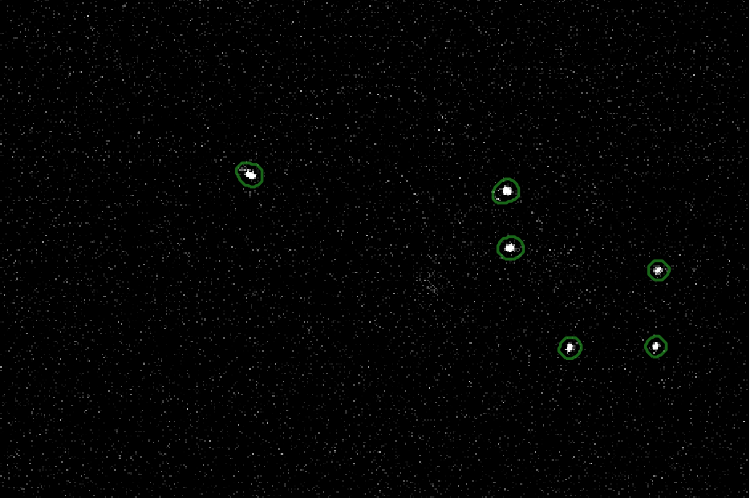} 
	\includegraphics[width=0.75\linewidth]{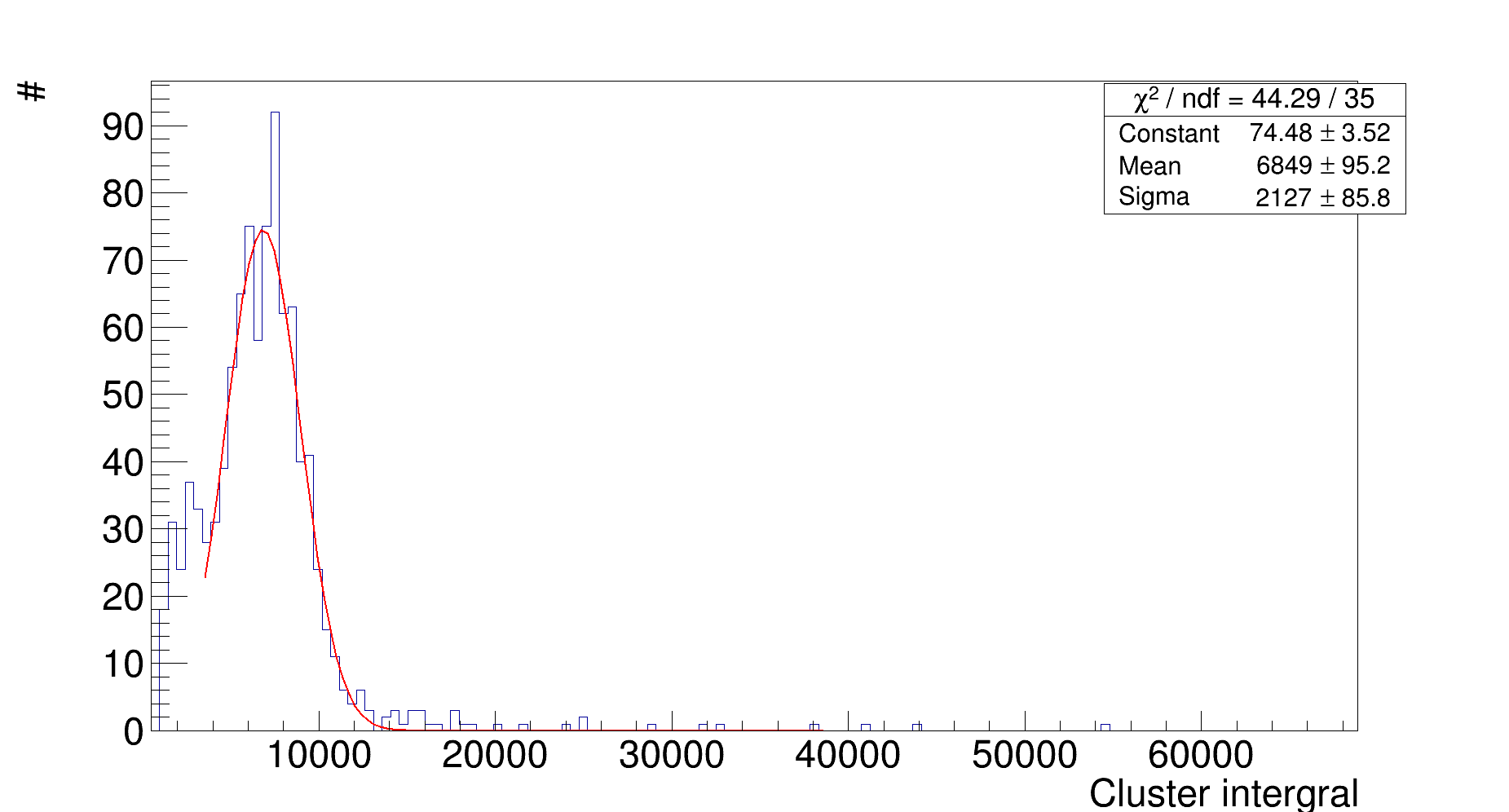}
	\caption{Example of $^{55}$Fe signals: on the top, an image acquired by the sCMOS camera in MANGO with a \emph{Tt} GEM configuration, He:CF$_4$ 60/40 gas mixture and 6 kV/cm induction field, where the $^{55}$Fe clusters are individually identified by the CYGNO reconstruction algorithm \cite{Baracchini:2020iwg,bib:cygnoIDBscan}; on the bottom, example of $^{55}$Fe photon spectrum with superimposed the Gaussian fit from the same configuration.}
	\label{fig:waveformGEM}
\end{figure}

Fig. \ref{fig:waveformGEM} top shows an example of 5.9 keV signals generated by the $^{55}$Fe X-ray  source  in MANGO as seen by the sCMOS camera. 

The images acquired are analysed with an iterative density based scanning algorithm (IDBSCAN) developed by the CYGNO collaboration \cite{Baracchini:2020iwg,bib:cygnoIDBscan} that searches for pixel clusters representing tracks with different energy deposition patterns, after having subtracted the camera noise pixel by pixel. An X-Y selection is applied to the clusters recognised by the algorithm removing events outside of a 3.45 $\times$ 3.45 cm$^2$ region at the center of the image, to avoid events at the border which may be compromised by drift field distortion, as confirmed by simulation in Sec. \ref{sec:maxwell}, and optical distortions due to the lens. 

The distance travelled in the gas by the primary electrons created by the interaction of the $^{55}$Fe X-rays in the gas target is of the order of $\mathcal{O}$(100) $\mu$m. The 2D X-Y projection of the signal imaged by the sCMOS camera hence appears round, being the shape dominated by diffusion and not by the original topology of the track. For this reason, the ratio of the minor over the major axis of an ellipse containing the cluster (defined as \emph{slimness}) is required to be larger than 0.7, to reject long straight tracks from cosmic rays or short curly tracks from natural radioactivity. 

For each of the found clusters satisfying the selection requirement described above, the energy deposited is calculated from the sum of the content of all the pixels belonging to the track (Integral), after noise subtraction. In addition, the dimension of the $^{55}$Fe round spot encodes the information of the diffusion suffered by the electron track from its production to the detection point. Due to the very small drift gap of 0.8 cm in MANGO and the value of transverse diffusion of about 100 $\frac{\mu m}{\sqrt{cm}}$ at the drift field of operation \cite{Amaro:2022gub}, the spot dimension is dominated by the contribution of the amplification stage rather than diffusion during drift. Therefore, the analysis of the $^{55}$Fe spot dimension provides important information on the intrinsic diffusion due to the GEMs employed and the choice of stacking. \\
\begin{figure}[!t] 
	\centering
	\includegraphics[width=0.75\linewidth]{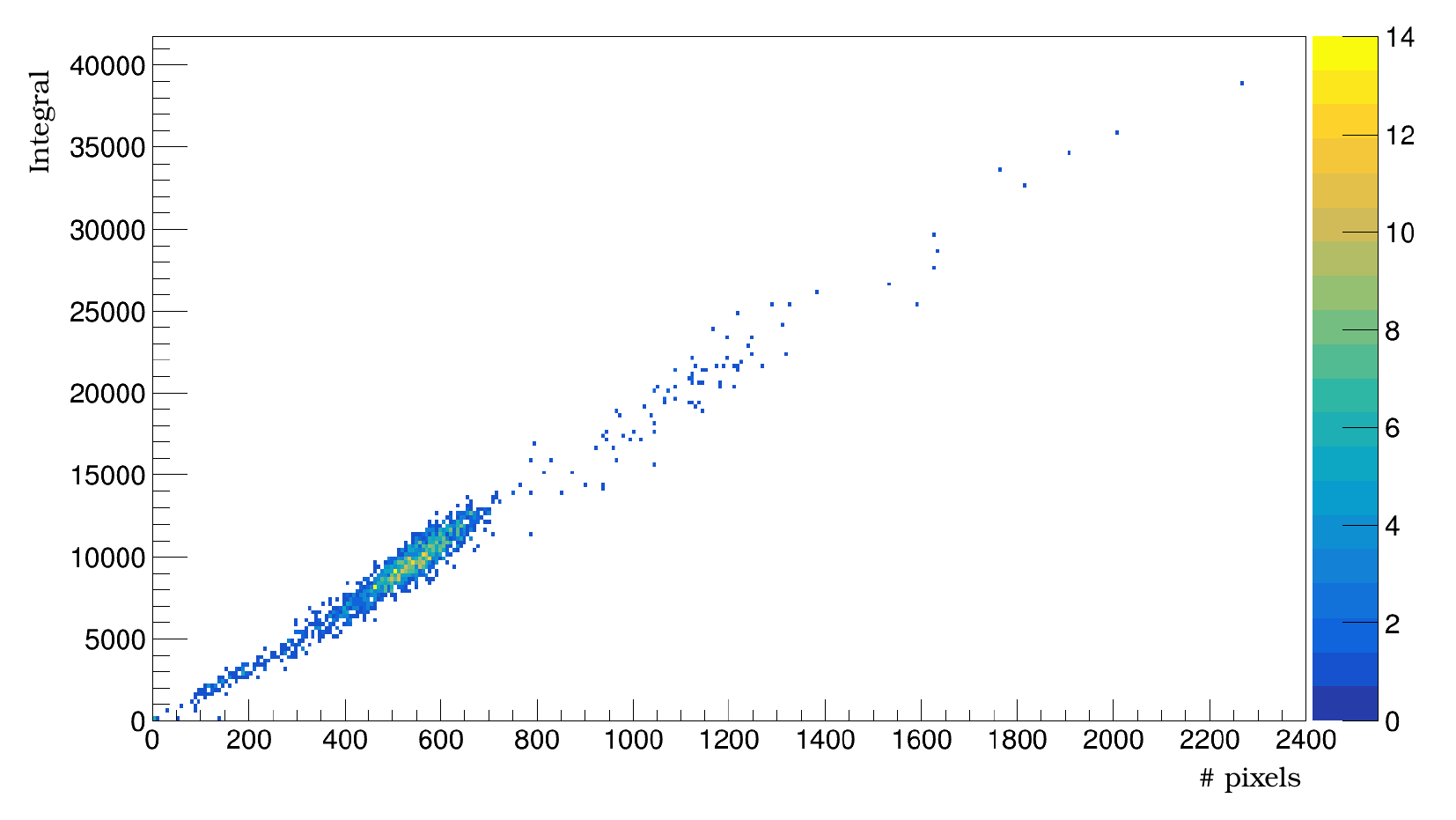}
	\includegraphics[width=0.75\linewidth]{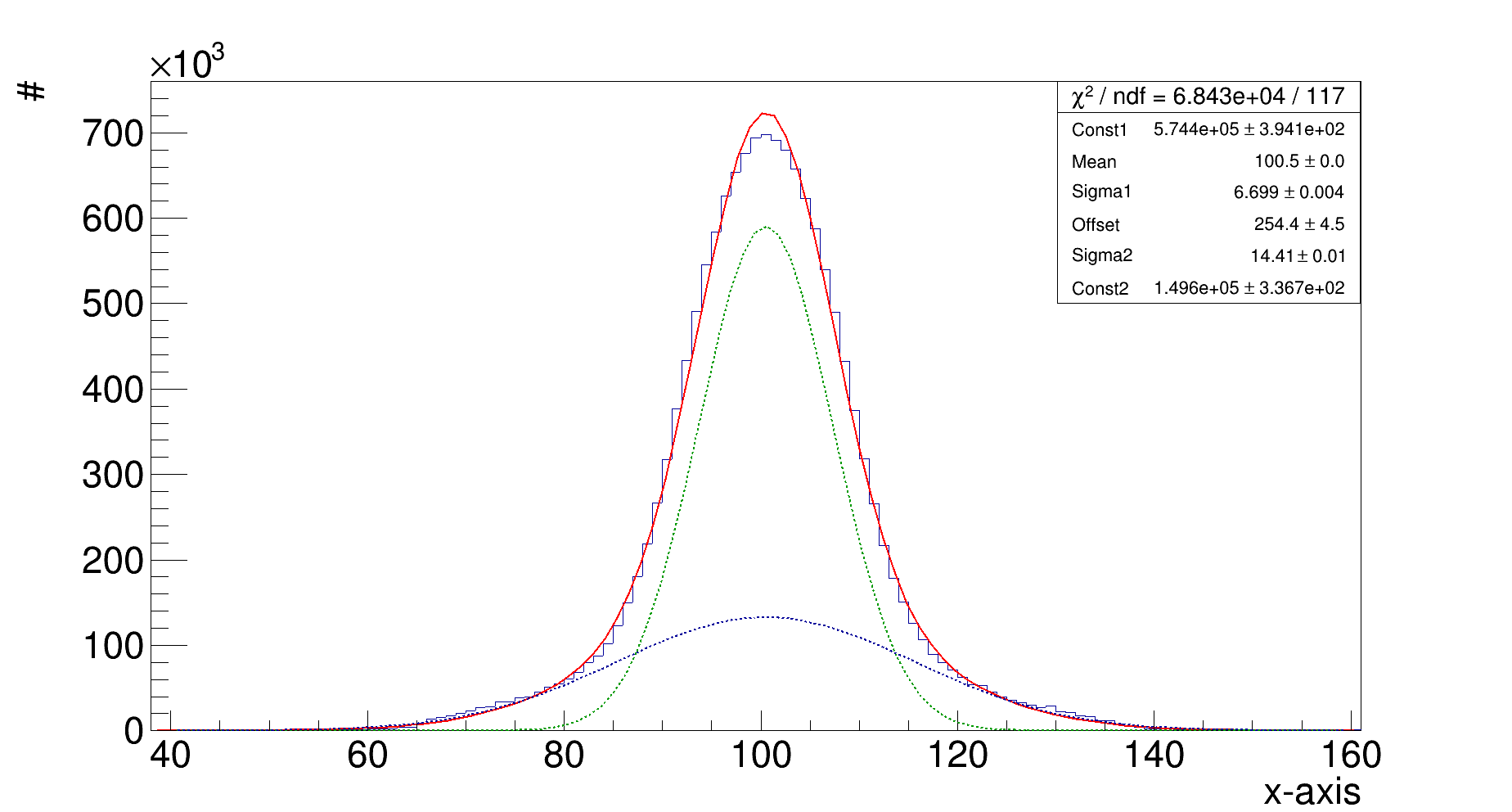}
	\caption{On the top, light integral of the selected $^{55}$Fe clusters versus the number of pixels included in the cluster by the reconstruction algorithm. On the bottom, X projection of the $^{55}$Fe centered clusters with a Double Gaussian fit superimposed in red. In green and blue the two separate Gaussian functions are displayed, with the green one being the \emph{primary} as described in the text. All axes of both graphs are in arbitrary count units.}
	\label{fig:nhits}
\end{figure}

\begin{figure*}[!t] 
	\centering
	\includegraphics[width=0.93\linewidth]{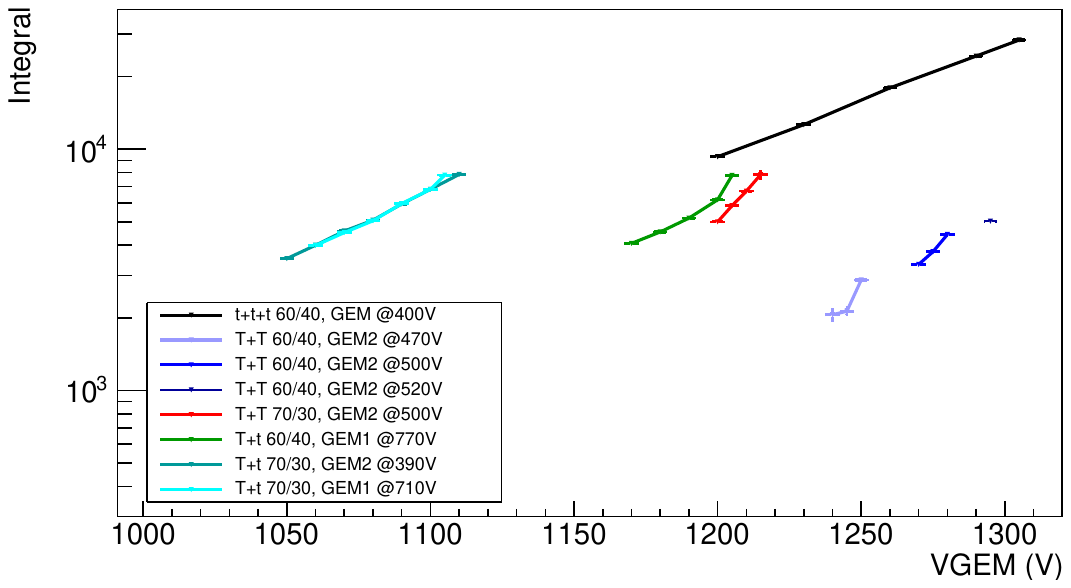}
	\caption{Gain scan summarizing plot. The light integral obtained by the $^{55}$Fe analysis are shown as a function of the total sum of the voltage applied across the GEMs. Different colours represent the various amplification and gas mixture combinations.}
	\label{fig:gain}
\end{figure*}
\begin{table*}[!t]
	\centering
	\begin{adjustbox}{max width=0.8\linewidth}
		\begin{tabular}{|c|c|c|c|c|}
			\hline
			
			\large{Ampl stage} & \large{Color on plot} & A'  & B' [1/V]  & Avg B' [1/V]    \\ \hline \hline
			ttt 60/40 & Black & -3.7 $\pm$ 0.3 & 0.0107 $\pm$ 0.0003 & 0.0107 $\pm$ 0.0003 \\ \hline
			Tt 60/40 & Green & -8 $\pm$ 1 & 0.0140 $\pm$ 0.0009 &  \\
			Tt 70/30 & Cyan & -7.3 $\pm$ 0.7 & 0.0147 $\pm$ 0.0006 & 0.0139 $\pm$  0.0004 \\ 
			Tt 70/30 & Dark Green & -5.8 $\pm$ 0.6 & 0.0133 $\pm$ 0.0005 &  \\  \hline
			TT 60/40  & Blue & -28 $\pm$ 10 &  0.029 $\pm$ 0.008 & \multirow{2}{*}{0.030 $\pm$ 0.004} \\ 
			TT 70/30 & Red & -27 $\pm$ 7 & 0.030 $\pm$ 0.006 &  \\ \hline
		\end{tabular}
	\end{adjustbox}
	\caption{Table summarising the results of the fit with Eq.\ref{eq:gainsimpl} to the data sets in Fig.\ref{fig:gain}. All the blue data sets were fitted together.}		
	\label{tab:gain}
\end{table*}
In this respect, it is important to notice how a simplistic definition of the spot dimension in terms of number of pixels identified as belonging to the track by the reconstruction code \cite{Baracchini:2020iwg,bib:cygnoIDBscan} would result in a biased determination of the diffusion. The light integral and the number of cluster pixels above threshold are in fact highly correlated, as can be seen in the top panel of Fig. \ref{fig:nhits}. In order to employ a variable independent from the track energy to properly evaluate the track diffusion, the distribution of the X and Y projections of the cluster is studied, since the shape is expected to be preserved even if the cluster becomes more luminous. With the goal of minimising any systematic effect and provide a robust diffusion estimation, the $^{55}$Fe spot cluster projection distributions are averaged after having aligned all their barycentres, where the barycentre is defined as the average x and y pixel coordinate weighted on the pixel intensity.
An example of this is shown in the bottom panel of Fig. \ref{fig:nhits}, with a Double Gaussian fit with a common mean superimposed. 

A Double Gaussian is used in order to properly include secondary tails in the projection distributions, that anyway never account for a fraction larger than 20\%. By reconstructing the $^{55}$Fe spots with an older version of the CYGNO collaboration reconstruction code and comparing the two, we verified that the observed tails in the projection distributions depend on the algorithm definition of a cluster boundary rather than by misalignment of the barycentres. The Gaussian function of the two which fits the core part of the projection distributions is called \textit{primary}. The diffusion is hence estimated by averaging the primary sigma, also called \textit{Sigma1}, of the X and the Y projection of the $^{55}$Fe spot.

The light output, the energy resolution and the diffusion of $^{55}$Fe-induced events are studied for each of the configurations illustrated in Tab. \ref{tab:datataking} both as a function of the voltage applied across the GEMs V$_{GEM}$ (i.e. the charge gain) and the intensity of the induction field after the last GEM amplification plane E$_{ind}$, and the results are illustrated in the following Sections.
\subsection{Light yield as a function of the charge gain}\label{subsubsec:gain}
In this section we present the study of the dependence of the light yield on the voltage applied across the GEM electrodes V$_{GEM}$ (that effectively defines the charge gain of the detector) without adding any field to the induction region. The values chosen for V$_{GEM}$ (and shown in Tab. \ref{tab:app}) depend, on the lower end, on the minimum voltage that allows the signal to be visible in the sCMOS images, and, on the higher end, on the voltages that are stable enough to keep the rate of sparks lower than 0.003 Hz.\\
The light spectrum of the selected events is modeled with a Gaussian function and the fitted mean is taken as the light integral. An example of a fitted $^{55}$Fe light spectrum is shown in Fig. \ref{fig:waveformGEM} bottom panel. The light integrals obtained by the $^{55}$Fe analysis for all the configuration of Tab. \ref{tab:datataking} are shown as a function of the total sum of the voltage applied across the GEMs in Fig. \ref{fig:gain}.

Since in this MANGO configuration the light is produced only in the electron avalanche amplification process happening within the GEMs, it is possible to interpret the results in Fig. \ref{fig:gain} in terms of detector charge gain. From the general description of the electron avalanche processes \cite{bib:tom,bib:Aoyama,bib:Williams,bib:Diethorn}, the reduced gain $\Gamma$ can be expressed as:\\
\begin{equation}
\label{eq:gain}
	\Gamma = \frac{ln(G)}{n_g pt} = A \left(\frac{V_{GEM}}{n_g pt}\right)^m exp\left(-B\left(\frac{n_g pt}{V_{GEM}}\right)^{1-m}\right)
\end{equation}
with $G$ the gain, the number of secondary electrons produced in the amplification per primary one, $n_g$ the number of GEMs used in the amplification stage, $p$ the gas pressure, $t$ thickness of the GEM, $V_{GEM}$ total voltage applied to the GEMs and $m$, $A$, $B$ free parameters. In particular, $m$ is constrained between 0 and 1 and depends on the gas. For gain scans that do not span over a large range of voltages, $m$ can be approximated to 1, resulting in the more widely used expression of the gain in a gas detector:
\begin{equation}
ln(G) = A'+B'V
\label{eq:gainsimpl}
\end{equation}  
Equation \ref{eq:gainsimpl} can then be used to fit all the data sets in Fig. \ref{fig:gain}, and the fit results are listed in Tab. \ref{tab:gain}.

The fit results show how each group of GEM stacking configuration (i.e. \emph{ttt}, \emph{Tt} and \emph{TT}) displays the same gain slope (B' parameter) independent from the gas mixture used. Conversely, the He to CF$_4$ ratio influences the voltage on the GEMs needed to attain the same gain, with larger helium content requiring lower voltages. The total light output achievable is of the same order of magnitude once the same amplification structure is put under examination, with the only exception being the \emph{TT} at 60/40. While the increase of helium results beneficial in terms of a lower amplification voltage, it significantly increases the frequency of sparks and cascade instabilities.\\
\begin{table}[!t]
	\centering
    \begin{adjustbox}{width= 1\linewidth}
	\begin{tabular}{|c|c|c|c|c|c|}
		\hline
		Config & Colour & $\eta$ $\left[\frac{1}{torr\cdot cm}\right]$ & $\beta$ $\left[\frac{1}{torr\cdot cm\cdot V}\right]$ \\ \hline \hline
		ttt 60/40 & Black & -0.36 $\pm$ 0.14 & 0.00106 $\pm$ 0.00011 \\ \hline
		Tt 60/40 & Green & -0.7 $\pm$ 0.2 & 0.0012 $\pm$ 0.0004 \\
		Tt 70/30 & Cyan & -0.6 $\pm$ 0.2 & 0.0012 $\pm$  0.0003 \\ 
		Tt 70/30 & Dark Green & -0.49 $\pm$ 0.19 & 0.0011 $\pm$ 0.0002  \\ \hline
		TT 60/40  & Blue & -1.6 $\pm$ 0.9 &  0.0017 $\pm$ 0.0007  \\ 
		TT 70/30 & Red & -1.6 $\pm$ 1.0 & 0.0018 $\pm$ 0.0006 \\ \hline
	\end{tabular}
    \end{adjustbox}
	\caption{Table summarising the results of the linear fit of the reduced light gain of all the configurations as a function of V$_{GEM}$, following Equation \ref{eq:gamm_vsvoltsigma}.}
	\label{tab:fit_gamma_vgem}
\end{table}
The larger gain and light output is achieved with the three thin GEM configuration \emph{ttt}, with integral values on average $\sim$ 3 times larger than \emph{Tt} and up to $\sim$ 5 than \emph{TT}. As expected, having more planes of GEMs grants higher amplification.
It is also interesting to notice that for the \emph{Tt} sets at 70/30, the scans are taken varying the voltage of the thick or the thin GEM alternatively. The light results of these sets are perfectly consistent as the points overlap each other nicely. This is consistent with the expectation of the gain dependence only on the total voltage applied across the GEMs, other than the stability of the detector during the data taking.\\
An interesting way to compare the different GEM amplification stacking options is by analysing their reduced gain $\Gamma$ (see Eq. \ref{eq:gain}) as a function of reduced field $\Sigma$, which is an approximation of the electric field inside the GEM holes normalised by some parameters of the experimental configuration. The latter is believed to effectively characterise the development of the electron avalanche and is defined as:
\begin{equation}
\label{eq:sigma}
    \Sigma = \frac{V_{GEM}}{n_g p t}
\end{equation}
As the assumption of the limited range of voltages utilised for the data taking of each scan is still valid, $m$ can be approximated to 1 and the reduced gain can be written as:
\begin{equation}
\label{eq:gamm_vsvoltsigma}
\Gamma = A_0+ B_0\Sigma= A_0+ \frac{B_0}{pn_gt}V_{GEM}.
\end{equation}
This is a simple mathematical recombination of the terms in play in order to highlight the dependence on the number of GEMs and the applied voltages. Indeed,  when comparing Eq. \ref{eq:gainsimpl} with Eq. \ref{eq:gamm_vsvoltsigma}, it is valid that $A'n_gpt=A_0$ and $B_0=B'$. The results of the fits of the data presented in Fig. \ref{fig:gain} with a function $$\Gamma = \eta + \beta V_{GEM},$$ are summarised in Tab. \ref{tab:fit_gamma_vgem}.

Once the terms proportional to $V_{GEM}$ are adjusted for, the number and thickness of GEMs, their fitted values result highly consistent among all the configurations and gas mixtures employed.\\

\subsection{Enhancing the light yield through the addition of strong induction field}
\label{subsubsec:el_light}
Given the importance of enhancing the light yield for optically readout TPCs as discussed in Sec. \ref{sec:intro} and the results of Sec. \ref{sec:maxwell}, the effect of the introduction of a strong electric field in the induction region is studied in details in this Section, expanding on the results present in \cite{bib:EL_cygno} and in Sec. \ref{sec:lemon}. Thus, an additional induction field E$_{ind}$ is applied to the induction region below the last GEM electrode and this effect is studied for all the GEM stacking configurations and gas mixtures illustrated in Tab. \ref{tab:datataking}. 
\begin{figure}[t] 
	\centering
	\includegraphics[width=1.1\linewidth]{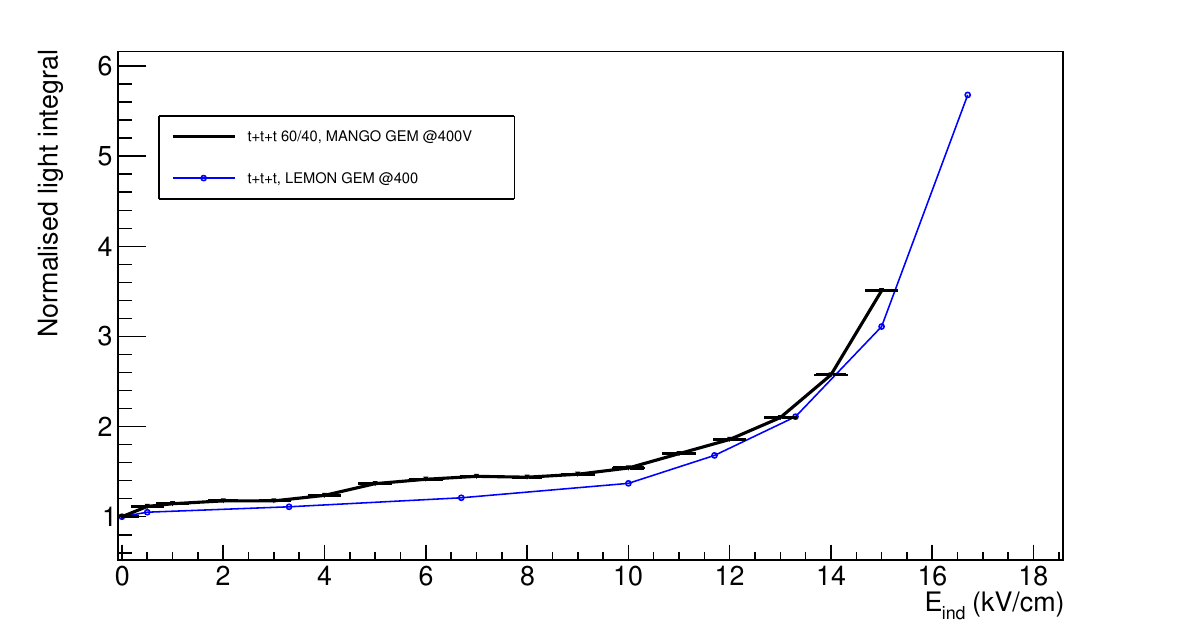}
	\caption{Relative increase of light integral for the \emph{ttt} configuration in MANGO and LEMOn.The two data sets are manifestly highly consistent with each other and with the measurements presented in \cite{bib:EL_cygno}, robustly confirming the results presented in Sec. \ref{subsec:mango_light}. }
	\label{fig:ellightttt}
\end{figure}

In Fig. \ref{fig:ellightttt} the relative increase in light yield with respect to the absence of induction field applied is shown in black for a \emph{ttt} configuration with He:CF$_4$ 60/40 and 400 V applied across each GEM as a function of the induction field E$_{ind}$. The relative light increase measured with the LEMOn detector and discussed in Sec. \ref{sec:lemon} is superimposed in blue. The two trends are strongly consistent with each other and show the same features, demonstrating that the light yield enhancement does not dependent on a single detector characteristics, but instead, it may be more likely associated to a physical phenomenon happening with the amplification structure under study. This last argument gains strength as we emphasise that the two detectors employ identical voltages applied to the GEMs, but with different absolute gain, since LEMOn is located at LNF at about 150 m a.s.l., while MANGO at the LNGS at 1000 m a.s.l., highlighting the independence of the relative light output growth from the absolute gain of the detector. The two detectors moreover employ different structures to apply the induction field E$_{ind}$ (a metallic mesh in MANGO and an ITO glass in LEMOn) demonstrating that, once the transparency of these structures is properly taken into account, the light yield amplification results independent from this feature.
\begin{figure}[t] 
	\centering
	\includegraphics[width=0.9\linewidth]{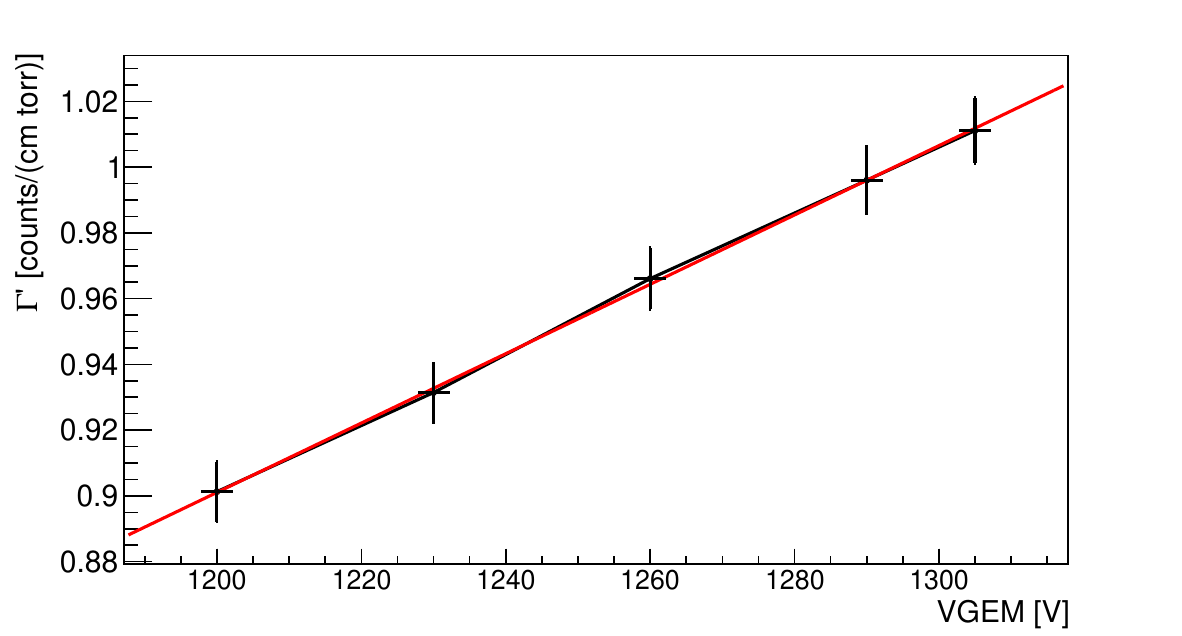}
	\includegraphics[width=0.9\linewidth]{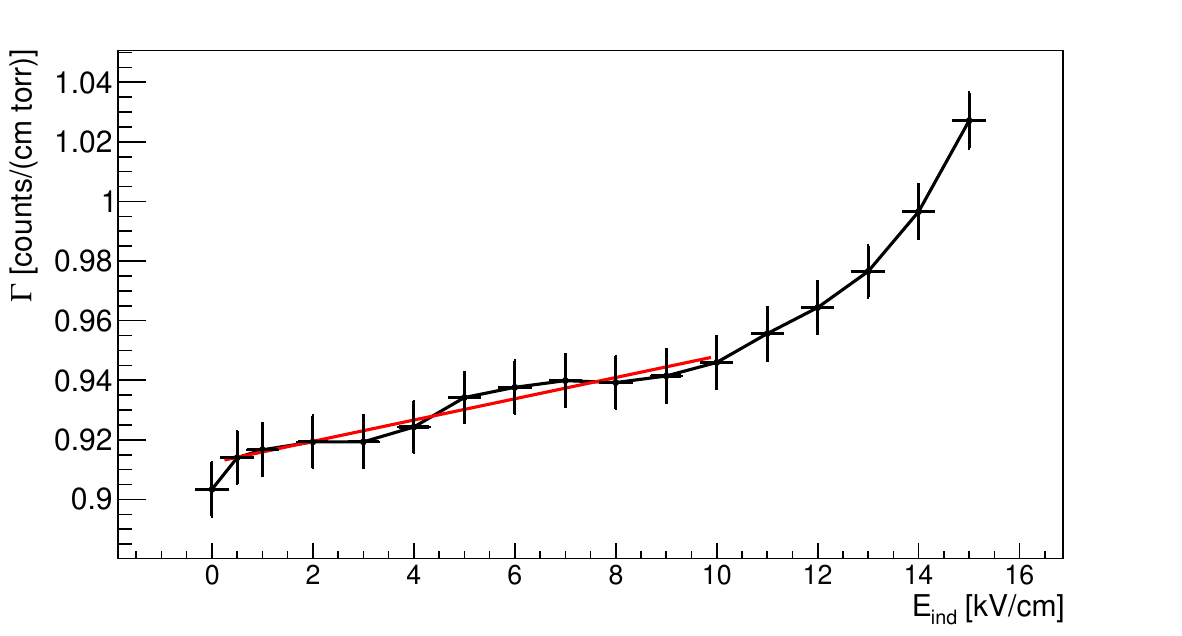}
	\caption{On the top, the reduced light gain as a function of V$_{GEM}$ with a linear fit superimposed. On the bottom, the reduced light gain is expressed as a function of $E_{ind}$ with a linear fit superimposed in the region blow 10 kV/cm.}
	\label{fig:fitEh_gain}
\end{figure}
The dependence of the light gain on the induction field is common to all the stacking configurations and gas mixtures studied, and can be split into three regions. Firstly, as soon as the field is turned on there is a boost in the light output of about 10\%. Afterwards, from 0.5 kV/cm  up to breaking point $E_b$ between 7 kV/cm and 10 kV/cm, the light grows linearly with E$_{ind}$, and beyond it, the light yield increase becomes exponential. This breaking point, where the increase changes from linear to exponential, is observed to depend on the gas mixture, being about  $E_b$ = 10 kV/cm for 60/40, and $E_b$ = 8 kV/cm for 70/30.  Despite the difference in the type of analysis of the two data sets (see Sec. \ref{sec:lemon}), this behaviour is consistently shared also between the MANGO and the LEMOn data. The features of each of these regions is discussed in detail in the following.

\paragraph{Region between 0 and 0.5 kV/cm} The enhancement generated by a field between 0 kV/cm and 0.5 kV/cm can be explained by the fact that typical MANGO (and LEMOn and all CYGNO prototypes) operation foresees the bottom of the last GEM electrode to be put to ground, to minimise the overall HV needed to be applied since no charge needs to be collected with an optical readout, as discussed in Sec. \ref{sec:lemon}. This implies that the field lines, typically showing an unordered closure on the lower electrode of the GEM, get straightened and align more systematically with this additional small electric field, resulting in a slight light yield increase. This hypothesis is confirmed by the simulation discussed in Sec. \ref{sec:maxwell}.
\begin{table}[!t]
	\centering
	\begin{adjustbox}{width=1\linewidth}
		\begin{tabular}{|c|c|c|c|c|}
			\hline
			Conf  & $\gamma$ $\left[ \frac{1}{torr\cdot cm}\right]$  & $\delta$ $\left[\frac{1}{torr\cdot kV}\right]$ &   Cond. C & $\alpha$  \\ \hline \hline
			ttt 60/40  & 0.912 $\pm$ 0.005 & 0.0036 $\pm$ 0.0010 & $\checkmark$ & 0.23 $\pm$ 0.06\\ \hline
			Tt 60/40  & 0.747 $\pm$ 0.009 & 0.0010 $\pm$ 0.0009 & $\checkmark$ & 0.05 $\pm$ 0.04 \\
			Tt 70/30  & 0.704 $\pm$ 0.007 & 0.0029 $\pm$ 0.0015  & $\checkmark$ & 0.14 $\pm$ 0.07 \\ \hline
			TT 60/40   & 0.48 $\pm$ 0.01 &  0.0025 $\pm$ 0.0013 & $\checkmark$ & 0.06 $\pm$ 0.03\\ 
			TT 70/30  & 0.505 $\pm$ 0.004 & 0.0016 $\pm$ 0.0010 & $\checkmark$ & 0.04 $\pm$ 0.03\\ \hline
		\end{tabular}
	\end{adjustbox}
	\caption{Table summarising the results of the linear fit of the reduced light gain as a function of the $E_{ind}$, when V$_{GEM}$ is fixed. Cond. C is marked with a $\checkmark$ if that condition is fulfilled.}
	\label{tab:fit_gamma_mesh}
\end{table}
\paragraph{Linear region between 0.5 kV/cm and E$_b$} In the region between 0.5 kV/cm and a breaking point $E_b$, the light yield increase appears linearly proportional to the raise in induction field E$_{ind}$. Starting from the arguments presented in Sec. \ref{subsubsec:gain} and the Maxwell simulations of Sec. \ref{sec:maxwell}, it is reasonable to believe that the induction field effectively enhances the reduced field $\Sigma$ inside the GEM holes with a linear relation. To include a contribution from the $E_{ind}$, an additional term to Eq. \ref{eq:gamm_vsvoltsigma} can be added as:

\begin{equation}
\label{eq:sigma_ext}
\Sigma = \frac{1}{p}\left(\frac{V_{GEM}}{n_gt}+\alpha E_{ind}\right) ,
\end{equation}
with $\alpha$ the coefficient of proportionality of $E_{ind}$. Therefore, the dependence of the reduced gain $\Gamma$ on the V$_{GEM}$ and $E_{ind}$ when m = 1 can be expressed as:
\begin{equation}
\label{eq:gamm_vsmesh}
\Gamma = A_0+ \frac{B_0}{pn_gt}V_{GEM}+\frac{B_0\alpha }{p}E_{ind}
\end{equation}
When $E_{ind}$ is zero, Eq. \ref{eq:gamm_vsvoltsigma} is recovered. On the contrary, if V$_{GEM}$ is fixed and the $E_{ind}$ is increased, a linear increase of $\Gamma$ is expected. Fig. \ref{fig:fitEh_gain} shows the reduced gain $\Gamma$ for \emph{ttt} configuration with He:CF$_4$ 60/40 on the top as a function of the $V_{GEM}$, with $E_{ind}=0$, and on the bottom as a function of $E_{ind}$ with $V_{GEM} = 400$, fitted with Eq. \ref{eq:gamm_vsmesh} up to $E_b$, display the linear dependence of the light yield on both quantities. 
\begin{figure*}[!t] 
	\centering
	\includegraphics[width=0.8\linewidth]{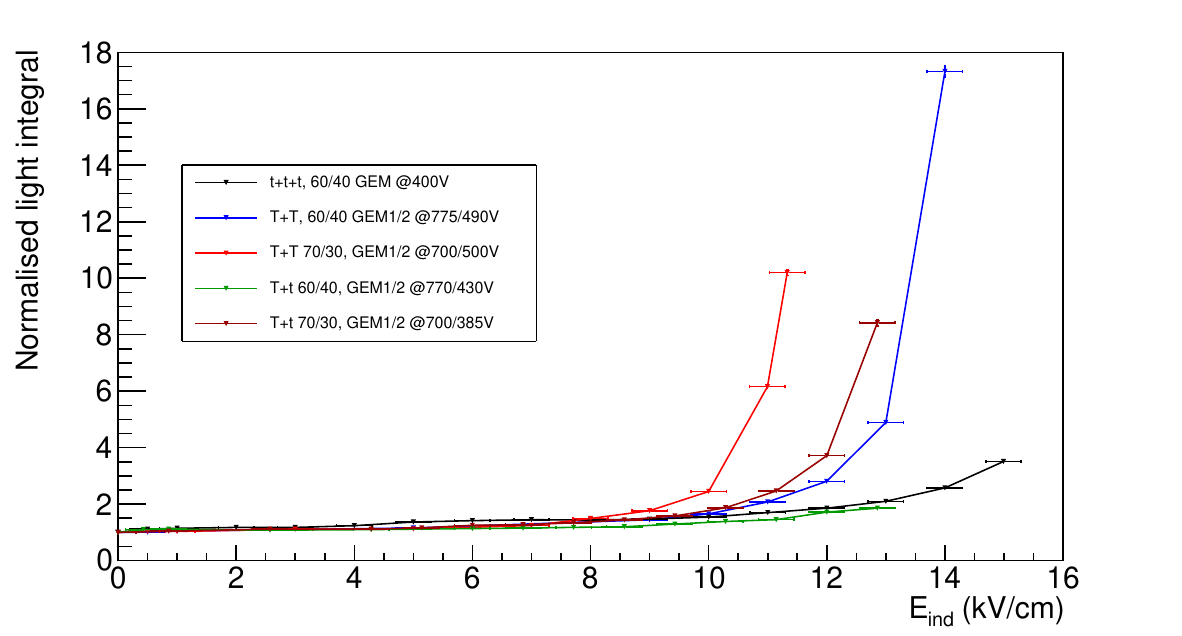}
	\caption{Relative increase of light output as a function of the induction field for all the GEMs stacking configurations studied with MANGO after the linear component is subtracted as described in the text.}
	\label{fig:elall_light}
\end{figure*}
\begin{table*}[!t]
	\centering
	\begin{adjustbox}{width=0.65\linewidth}
		\begin{tabular}{|c|c|c|c|c|c|}
			\hline
			
			\large{Config} & $a$ & $b$ & $c$ [cm/kV] & $d$ & \small{E$_{b}$ [kV/cm]} \\ \hline  \hline
			ttt 60/40 & 0.99 $\pm$ 0.02 & 0.04 $\pm$ 0.02 & 0.8 $\pm$ 0.1 & 8.2 $\pm$ 0.6 & $9.8$  $\pm$  $1.5$ \\ 
			TT 60/40 & 0.99 $\pm$ 0.03 & 0.09 $\pm$ 0.03 & 1.1 $\pm$ 0.1 & 10 $\pm$ 1 &$9.5$  $\pm$  $1.2$ \\ 
			Tt 60/40 & 1.00 $\pm$ 0.02 & 0.08 $\pm$ 0.03 & 0.8 $\pm$ 0.2 & 8 $\pm$ 1 & $10$  $\pm$  $2$ \\ 
			Avg. 60/40 &  &  &  &  & $9.7$ $\pm$ $0.8$ \\ \hline 
			TT 70/30 & 1.00 $\pm$ 0.02 & 0.3 $\pm$ 0.2 & 1.2 $\pm$ 0.1 & 10 $\pm$ 2 & $8.5$  $\pm$  $1.3$ \\ 
			Tt 70/30 & 0.99 $\pm$ 0.02 & 0.17 $\pm$ 0.08 & 0.86 $\pm$ 0.03 & 7.6 $\pm$ 0.9 & $8.8$  $\pm$  $0.9$ \\ 
			Avg. 70/30  &  &  &  & & $8.7$ $\pm$ $0.7$ \\ \hline  
		\end{tabular}
	\end{adjustbox}
	\caption{Table summarising the result of the fit with Eq. \ref{eq:fitel_light} to the data of Fig. \ref{fig:elall_light}, where the field E$_{b}$ represents the value at which the exponential light increase growth starts.}
	\label{tab:elbreak}
\end{table*}

We can hence perform a linear fit to the reduced gain $\Gamma$ data as a function of $E_{ind}$ with the function
\begin{equation}\label{eq:gamma_mesh}
    \Gamma = \gamma + \delta E_{ind}
\end{equation}
between 0.5 kV/cm and $E_b$ for each configuration under study, to verify our assumption and measure the dependence on $E_{ind}$ of the light yield enhancement.
If the assumption that $E_{ind}$ contributes linearly to increase the effective field inside the GEMs holes is correct, the fitted $\gamma$ term has to be equal to $A+\frac{B}{pn_gt}V_{GEM,0}$, where V$_{GEM,0}$ is the sum of the voltages applied to the GEMs. We define this \emph{Condition C}. The first order term of the linear fit, $\delta$, allows to estimate the parameter $\alpha$, that defines the proportionality of the light increase to $E_{ind}$. Tab. \ref{tab:fit_gamma_mesh} shows the result of the fit with Eq.\ref{eq:gamma_mesh} to all the configurations considered in this study, together with \emph{Condition C} and the $\alpha$ proportionality parameter. \emph{Condition C} is always verified, demonstrating a clear comprehension of the contribution of $E_{ind}$ to the linear part of light yield increase and consistency with the Maxwell simulations.

\paragraph{Exponential region above E$_b$} The light yield enhancement beyond $E_b$ clearly stops to be consistent with a linear increase, allowing to conclude that above this value a new phenomenon comes into play. In order to properly study this additional feature, the fitted function from Eq. \ref{eq:gamma_mesh} is subtracted from the data in the entire range for each configuration for the following discussion. The relative light yield increase resulting after this subtraction is shown in Fig. \ref{fig:elall_light}.

All the curves display very similar behaviours, including the three different light enhancement regions discussed above, where the differences between gas mixtures and amplification structures is highlighted by the different breaking points $E_b$ and different exponential rise. 
The subtraction of the linear $E_{ind}$ enhancement enables direct comparison between the different configurations in the entire range of the data by employing a modified expression to describe the increase of the light yield as:
\begin{equation}
\label{eq:fitel_light}
a+b \cdot e^{cE_{ind}-d}
\end{equation}
where $a$ represents the normalisation with respect to operation with null induction field E$_{ind}$, $b$ is the intensity of the exponential component, $c$ determines how steeply the exponential grows, and $d$ is a shift in E$_{ind}$ field. The ratio between $d$ and $c$ returns the field value E$_{b}$ where the exponential growth starts, highly consistent with assumption used to evaluate the linear growth in the above discussion. 
Tab. \ref{tab:elbreak} summarises the parameters obtained by fitting the different data sets with Eq. \ref{eq:fitel_light}.

The results obtained show how, in the gas mixture tested, the larger the helium concentration in the mixture, the smaller the E$_{ind}$ value required to initiate the exponential growth of light. This is consistent with what was observed in Sec. \ref{subsubsec:gain} in terms of lower voltage across the GEM needed to start the electron avalanche responsible for the charge gain for higher helium fractions. For what concerns the phenomenological origin of this exponential rise, it can be related to the Maxwell simulation results (Sec. \ref{sec:maxwell}). Given the large E$_{ind}$ employed, the zone below the GEM holes of the last amplification GEM provides a wide and intense enough field to account for a strong increase in light yield. In addition, the exponential steepness parameter $c$ of the \emph{TT} configurations appears to represent a stronger boost of the light yield with respect to the other configurations, which are characterised by a $t$ GEM  next to the induction gap. This is coherent with the expectations driven after the Maxwell simulation as the E$_{ind}$ generates below the $T$ GEM a wider region where light production can occur.

\subsection{Energy resolution}
\label{subsec:res}

\begin{figure}[!t] 
	\centering
	\includegraphics[width=1.1\linewidth]{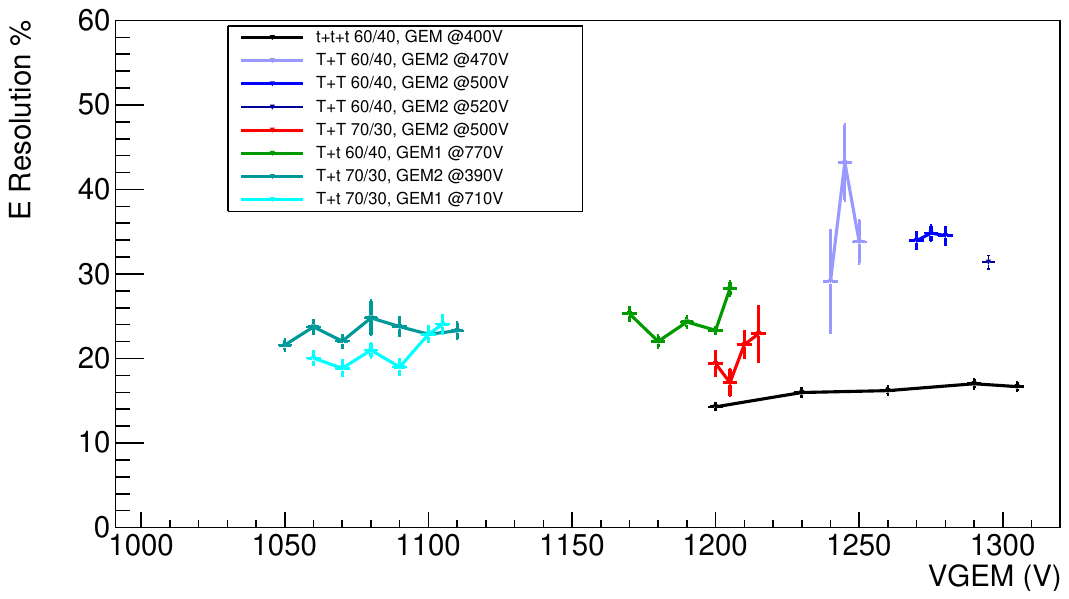}
	\caption{Energy resolution for the different amplification stages as a function of the sum of the  voltages applied to the GEMs with a null induction field. Different colours represent the various amplification and gas mixture combinations.}
	\label{fig:gaineres}
\end{figure}

The energy resolution at 5.9 keV is evaluated as the ratio of the sigma over the mean value of the Gaussian fit of the $^{55}$Fe spectrum and presented in this Section. The results as a function of the different GEM configurations and voltages with a null induction field applied are shown in Fig. \ref{fig:gaineres}. The energy resolution appears to strongly depend on the GEMs configuration used, spanning from 15\% up to 35\%, but seems unaffected by the gas mixture used. 

\begin{figure}[!t] 
	\centering
	\includegraphics[width=1.1\linewidth]{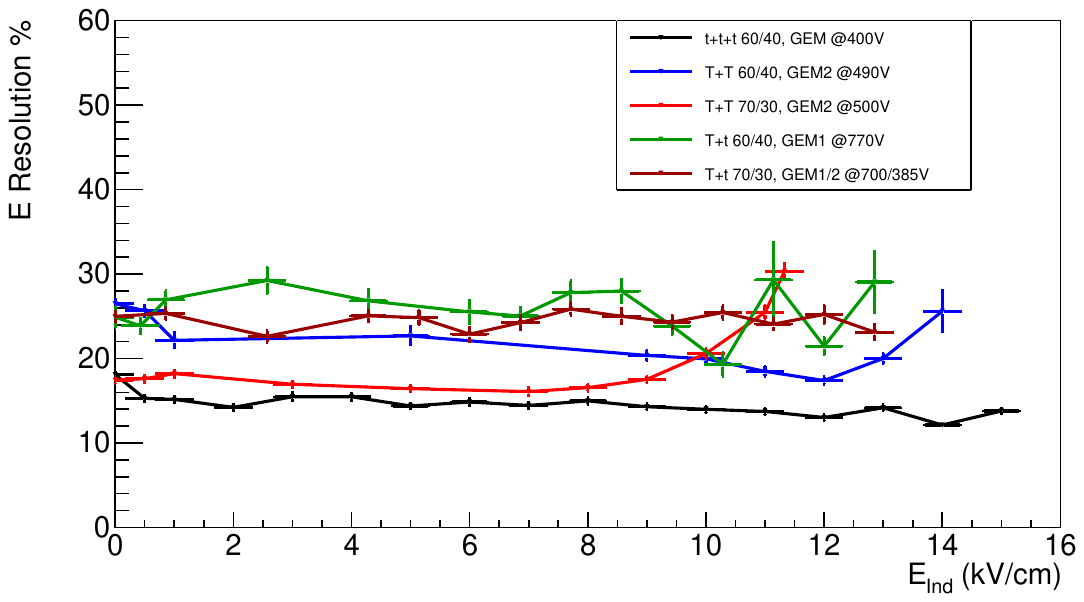}
	\caption{Energy resolution for the data sets with applied induction fields E$_{ind}$ as a function of E$_{ind}$.}
	\label{fig:el_res}
\end{figure}

The energy resolution for the data sets as a function of the induction fields E$_{ind}$ is shown in Fig. \ref{fig:el_res}.

In this case, the energy resolution appears to remain constant independently from the light yield increase induced by the E$_{ind}$ field. This is in line with the hypothesis underlying this entire paper, that it is possible to amplify the light output of a gas detector without relevant additional charge gain. In fact, if the light yield enhancement were due to additional electron avalanches generated in the induction gap within a reduced electric field much lower than those present in the GEM holes, the gain fluctuations would increase resulting in a worsening of the energy resolution \cite{bib:tom,bib:Alkhazov,bib:Schlumbohm}. This data further supports the idea that the enhancement of light is related to the findings of the Maxwell simulations presented in this paper. The exceptions to what has just been described are the \emph{TT} configurations at high fields. In this case the energy resolution is noticeably worsening with strong induction field, above the $E_b$, following again an exponential growth.
\subsection{Diffusion within the amplification stage}
\label{subsec:spot}
\begin{figure}[!t] 
	\centering
	\includegraphics[width=1.1\linewidth]{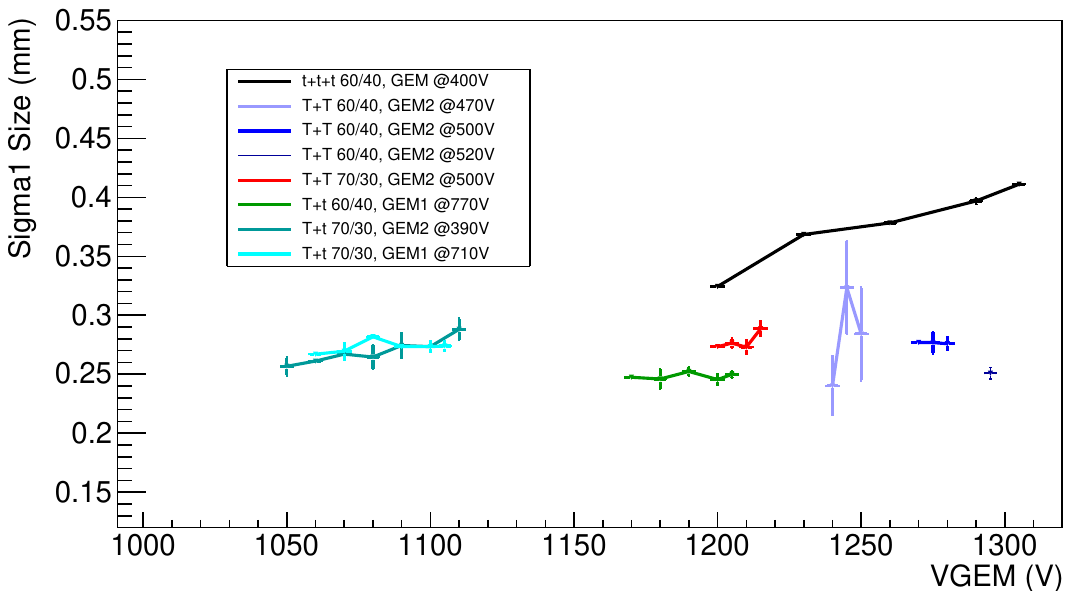}
	\caption{Amplification stage diffusion as a function of the sum of the voltages across the GEMs. Different colours represent the various amplification and gas mixture combinations.}
	\label{fig:diffsig1}
\end{figure}

As illustrated in Sec. \ref{subsec:mango_light}, a proper analysis of the $^{55}$Fe spot size allows us to measure the contribution to the original track diffusion caused by the amplification stage. Fig. \ref{fig:diffsig1} shows the primary sigma (averaged from the X and Y projections) as a function of the voltage applied to the GEMs for the different setups of amplification. Most of the double GEM structures are performing better than the triple GEM one, supporting the assumption that each stage of amplification contributes with an independent term to the overall diffusion.
The \emph{Tt} stacking configurations perform better than the \emph{TT} within the same gas mixture, in line with the expectation that the granularity of the GEM closer to the sCMOS sensor sets the maximum achievable space resolution if larger than the camera pixels. Having a \emph{t} GEM a pitch of 140 $\mu$m with compared to 350 $\mu$m of the \emph{T}, the effective pixel size of the MANGO setup of 49 $\times$ 49 $\mu$m$^2$ results sensitive to this feature. \\
Another interesting feature is that, while the triple GEM configuration diffusion linearly depends on the voltage applied to the GEM, the double ones display a much less significant (in some cases nonexistent) increase. More recent data acquired with MANGO (but not discussed in this paper) seems to suggest that the very large charge gain achieved with \emph{ttt}, coupled with the very small holes dimensions of these GEMs, are generating space charge effects that could be further worsening the diffusion in the amplification stage. This hypothesis is under study by the CYGNO collaboration and will be the subject of an upcoming paper.
The smaller diffusion measured with the \emph{Tt} configuration with respect to the \emph{TT} further demonstrates that the method developed to evaluate the diffusion within the GEMs and illustrated in Sec. \ref{subsec:mango_light} is independent of the light yield, having the first configuration a larger light output than the second.

\begin{figure}[!t] 
	\centering
	\includegraphics[width=1.1\linewidth]{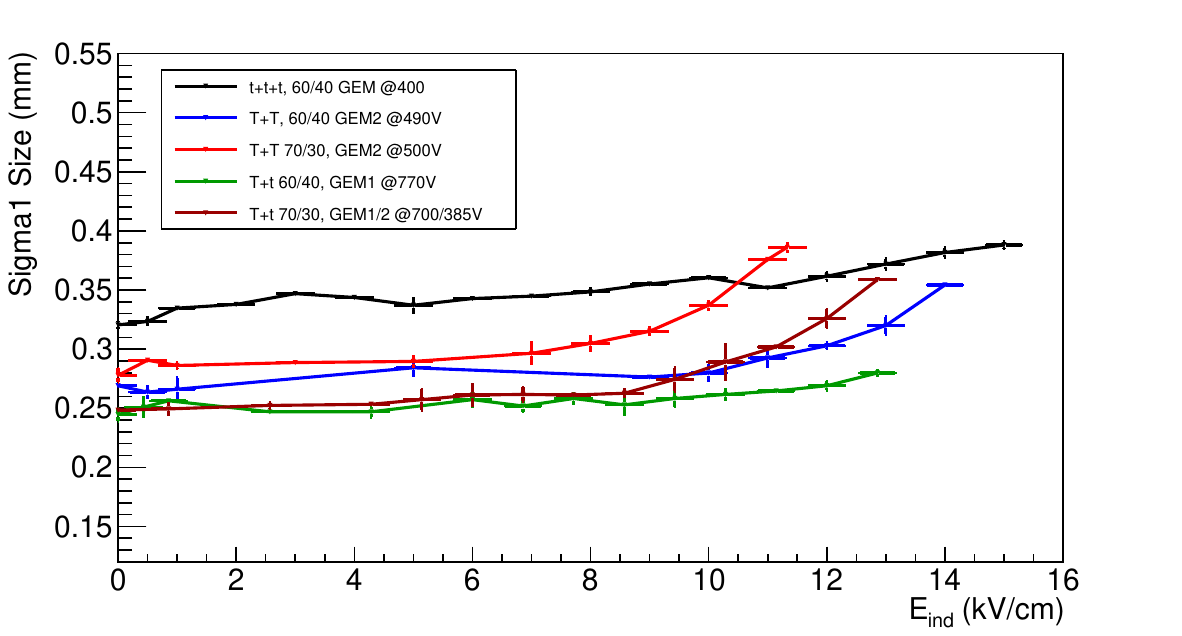}
	\caption{Amplification stage diffusion as a function of the $E_{ind}$  induction field.}
	\label{fig:el_size}
\end{figure}

\begin{figure}[!t] 
	\centering
	\subfloat[][]
	{\includegraphics[height=3.5 cm]{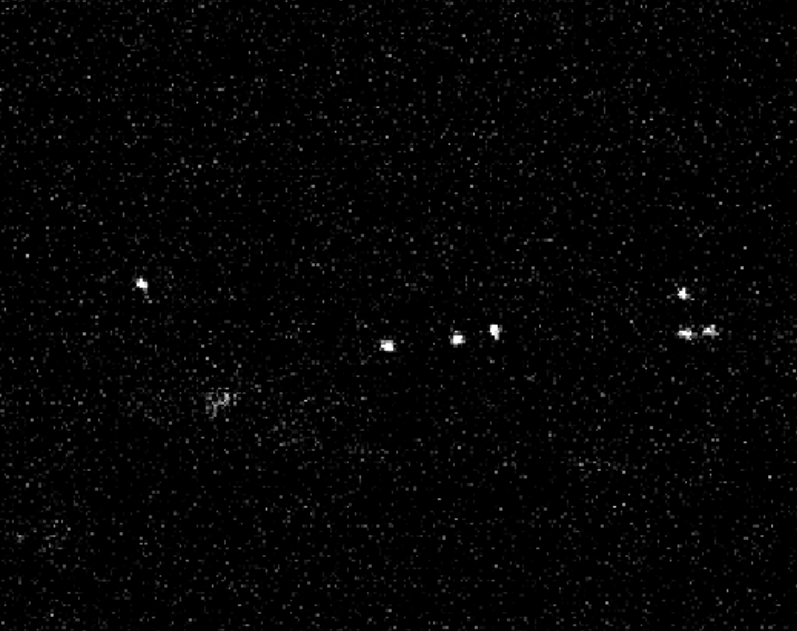}} \quad
	\subfloat[][]
	{\includegraphics[height=3.5 cm]{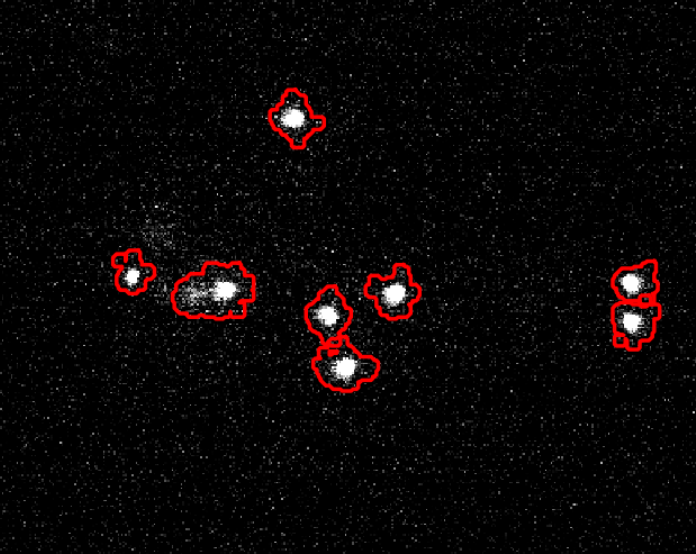}}
	\caption{Raw images of the $^{55}$Fe data taking with the \emph{TT} amplification structure and 60/40 of He:CF$_4$ gas mixture. On the top a picture with the E$_{ind}= 0$ kV/cm, while on the bottom the field is 11 kV/cm.}
	\label{fig:TTlargespots}
\end{figure}

The diffusion at the amplification stage (where now this includes the induction gap) is further investigated as a function of the $E_{ind}$ field and shown in Fig. \ref{fig:el_size}. These results corroborate the assumption that no large amount of charge is generated by electrons travelling in the induction gap, which would otherwise result also in a large spread of the additional light generated. Since the sCMOS camera is focused on the last GEM electrode (and could not anyway be focused on a volume, but only on a plane), the overall final effect results in a modest blur only slightly affecting the $^{55}$Fe spot size. Indeed, the \emph{ttt} diffusion at the maximum applied voltage on the GEMs (light increase of a factor 3.0 with respect to 1200 V) is about 6\% larger than the diffusion at the maximum induction field E$_{ind}$ (light increase of a factor 3.5 with respect to null induction field at 1200 V).
The data show the same type of trend as the energy resolution and light yield, with a general change in behaviour after the E$_b$ fields in Tab. \ref{tab:elbreak} are applied. It can be noted that the \emph{TT} configurations are again marked by most increase in the diffusion among the tested amplification structures.  The increase in dimension is visible directly on the raw images as the example of the \emph{TT} 60/40 in Fig. \ref{fig:TTlargespots}.
The \emph{ttt} and \emph{Tt} spot size dimensions are generally less affected by the increment in induction field, with a growth of less than 20\% at the largest  E$_{ind}$ values tested. As the latter configurations have in common a $t$ GEM closer to the induction gap, this finding suggests that this phenomenon is affecting the two kinds of GEM differently.

\section{Discussion}
\label{sec:disc}
\begin{table*}[!t]
	\centering
	\begin{adjustbox}{max width=0.7\textwidth}
		\begin{tabular}{|c|c|c|c|c|}
			\hline
			\multicolumn{2}{|c|}{} & Integral & E res (\%) & Diff [$\mu$m] \\
			\hline \hline
			\multirow{3}{*}{\textit{ttt}} & min &  9510 $\pm$ 40 & 16.0 $\pm$ 0.3 & 320 $\pm$ 4\\
			& max V$_{GEM}$ & 28400 $\pm$ 110 & 16.6 $\pm$ 0.3 & 412 $\pm$ 5\\
			& max $E_{ind}$ & 33500 $\pm$ 140 & 13.8 $\pm$ 0.3 & 388 $\pm$ 5\\ \hline
			\multirow{3}{*}{\textit{TT}} & min &  3410 $\pm$ 20 & 28.0 $\pm$ 1.5 & 260 $\pm$ 3\\
			& max V$_{GEM}$ &  5090 $\pm$ 30 & 31.0 $\pm$ 0.6 & 255 $\pm$ 3\\
			&  max $E_{ind}$ &  58800 $\pm$ 300 & 25.7 $\pm$ 0.5 & 356 $\pm$ 5\\ \hline
			\multirow{3}{*}{\textit{Tt}} & min & 4600 $\pm$ 30 & 25.2 $\pm$ 0.5 & 245 $\pm$ 3\\
			& max V$_{GEM}$ &  7700 $\pm$ 40 & 27.8 $\pm$ 0.5 & 245 $\pm$ 3\\
			& max $E_{ind}$ &  11800 $\pm$ 50 & 26.8 $\pm$ 0.5 & 280 $\pm$ 4\\ \hline
		\end{tabular}
	\end{adjustbox}
	\caption{Summary for the three configurations at 60/40 (\textit{ttt}, \textit{Tt} and \textit{TT}) of the \textit{integral}, the energy resolution and the intrinsic diffusion in three scenarios: the minimum voltage applied to the GEMs with no induction field ("min"), the maximum voltages applied to the GEMs with no induction field ("max V$_{GEM}$"), and finally the minimum voltage applied to the GEMs with the maximum induction field ("max $E_{ind}$")}
	\label{tab:check2}
\end{table*}
The combination of Maxwell simulations (Sec. \ref{sec:maxwell}) with the results obtained in Sec. \ref{sec:lemon} and in Sec. \ref{subsec:mango_light} suggests that when the induction field is raised beyond E$_b$, the reduced field in the bottom region of the GEM hole and few micrometers below it is high enough to produce a simultaneous (but not directly proportional) amplification of light and charge, effectively enlarging and shifting the actual amplification region of this configuration. In the simulated $T$ GEM in Sec. \ref{sec:maxwell}, $E_{ind}$ appears to generate a larger region of additional amplification, with electric fields closer to the ones in the centre of the GEM hole. Once a single gas mixture is considered, this is consistent with the larger light output of the \textit{TT} configuration visible in Fig. \ref{fig:elall_light} and characterised by a larger $c$ parameter in Tab. \ref{tab:elbreak}, which represented the slope of the exponential in the relative light increase as a function of $E_{ind}$. Moreover, the larger size is also compatible with the faster degradation of the signal due to diffusion for the \textit{TT} configurations observed in Fig. \ref{fig:el_size} with respect to the ones with a $t$ on the bottom.
This result is suggesting that the innovative amplification strategy illustrated in this study effectively enhances the light amplification potentialities of the region below the GEM holes, while minimising the additional simultaneous charge production.\\

In this study different combinations of GEMs and a varying amount of helium in the gas mixture were analysed with respect to their light yield properties, energy resolution and diffusion. The 70/30 mixture results in higher probability of discharges which makes the detector unstable and prone to failure while not providing any particular improvement with respect to the standard 60/40. When only the voltage across the GEM is increased, the \emph{ttt} configuration returns the better energy resolution and light yield as expected, but at the cost of a larger diffusion, which also depended on the voltage applied. The \emph{Tt} configuration has a lower light yield, around 3 times less, and guarantees a lower diffusion, with a reduction of 30\%. The \emph{TT} behaves in the middle for what concerns the diffusion, but possesses a light yield about 5 times lower than the \textit{ttt}.\\
\begin{table*}[!t]
	\centering
	\begin{adjustbox}{max width=0.7\textwidth}
		\begin{tabular}{|c|c|c|c|c|c|}
			\hline
			\multicolumn{2}{|c|}{} & $E_{ind}$ [kV/cm] & Integral & E res (\%) & Diff [$\mu$m] \\
			\hline \hline
			\multirow{3}{*}{$0$} & \textit{ttt} & 0 $\pm$ 0 & 9510 $\pm$ 40 & 16.0 $\pm$ 0.3 & 320 $\pm$ 4\\
			& \textit{TT} & 12 $\pm$ 0.3 & 9420 $\pm$ 40 & 17.4 $\pm$ 0.4 & 302 $\pm$ 4\\
			& \textit{Tt} & 11.1 $\pm$ 0.3 & 9360 $\pm$ 40 & 27 $\pm$ 0.5 & 264 $\pm$ 3\\ \hline
			\multirow{3}{*}{$1$} & \textit{ttt} & 3 $\pm$ 0.3 & 11300 $\pm$ 50 & 15.5 $\pm$ 0.3 & 347 $\pm$ 5\\
			& \textit{TT} & 12.3 $\pm$ 0.4 & 11300 $\pm$ 50 & 17.9 $\pm$ 0.4 & 307 $\pm$ 4\\
			& \textit{Tt} & 12.3 $\pm$ 0.4 & 11300 $\pm$ 50 & 25.0 $\pm$ 0.5 & 273 $\pm$ 4\\ \hline
			\multirow{3}{*}{$2$} & \textit{ttt} & 15 $\pm$ 0.3 & 33500 $\pm$ 140 & 13.8 $\pm$ 0.3 & 388 $\pm$ 5\\
			& \textit{TT} & 14 $\pm$ 0.3 & 58800 $\pm$ 300 & 25.7 $\pm$ 0.5 & 356 $\pm$ 5\\
			& \textit{Tt} & 12.8 $\pm$ 0.2 & 11830 $\pm$ 50 & 26.8 $\pm$ 0.5 & 280 $\pm$ 4\\ \hline
		\end{tabular}
	\end{adjustbox}
	\caption{Summary, for each configuration, of the induction field, the integral, the energy resolution and intrinsic diffusion in three different scenarios. Scenario "0" refers to when the three configurations have the same light output and the \textit{ttt} has $E_{ind}=0$. Scenario "1" refers to when the three configurations have the highest light output equal to each other. Finally, scenario "2" refers to when the maximum induction field is applied to each configuration.}
	\label{tab:check1}
\end{table*}
The introduction of a strong electric field  below the last GEM amplification plane enhances the light gain without affecting heavily the diffusion and the energy resolution. This innovative way of utilising the GEM with this specific gas mixture allows to improve the performances of the diverse GEM stacks employed. Tab. \ref{tab:check2} summarises for the three configurations at 60/40 (\textit{ttt}, \textit{Tt} and \textit{TT}) the \textit{integral} (proportional to the light yield), the energy resolution and the intrinsic diffusion in three scenarios: the minimum voltage applied to the GEMs with no induction field ("min"), the maximum voltages applied to the GEMs with no induction field ("max V$_{GEM}$"), and finally the minimum voltage applied to the GEMs with the maximum induction field ("max $E_{ind}$"). It can be noted that the addition of the induction field always allows to reach a light output larger than what is possible employing the GEMs in the standard way. Moreover, the larger light yield is accompanied by a similar, if not better energy resolution. For the \textit{ttt} configuration the intrinsic diffusion is also improved when the induction field is employed in place of the increase of the voltage applied to the GEMs. Conversely, for the \textit{TT} and \textit{Tt} ones, the diffusion worsens. This can be explained by the fact that the intrinsic diffusion of the \textit{ttt} was shown to be dependent on the applied voltage on the GEMs, differently from the \textit{TT} and \textit{Tt} cases (see Fig. \ref{fig:el_res}).\\
The performances of the different stacking options in the 60/40 gas mixture can be compared among each other to find the best solution. Tab. \ref{tab:check1} summarises, for each configuration, the induction field, the integral, the energy resolution and intrinsic diffusion in three different scenarios. Scenario $0$ refers to when the three configurations have the same light output and the \textit{ttt} has $E_{ind}=0$. Scenario $1$ refers to when the three configurations have the highest light output equal to each other. Finally, scenario $2$ refers to when the maximum induction field is applied to each configuration.\\
Scenarios $0$ and $1$ represent similar setups for which it is possible to demonstrate that the introduction of the induction field allows the 2 GEM stacks to attain a light output identical to the 3 GEM stack with similar energy resolution (only for the \textit{TT}) and reduced intrinsic diffusion.\\
Scenario $2$ directly compares the maximum light output achievable by the different configurations. The largest area generated under the $T$ GEMs thanks to the strong induction field permits to reach light gain dramatically high, maintaining a smaller intrinsic diffusion than the standard \textit{ttt}. The light output of \textit{TT} exceeds by a factor 2 the maximum achievable with standard operation of the  \textit{ttt} GEM stack, keeping the energy resolution at 5.9 keV below 30\% and intrinsic diffusion around 350 $\mu$m. \\
The analyses performed stress that depending on the experimental need each configuration excels in one of the variable studied. The \textit{ttt} has an excellent energy resolution thanks to the high gain and reduced field intensity, the \textit{Tt} always has the smallest intrinsic diffusion well below 300 $\mu$m, and finally the \textit{TT} allows to reach the largest light yields. In the context of a directional DM experiment, the largest impact on the sensitivity comes from the energy threshold \cite{Billard_2012}. CYGNO's expected energy threshold of 0.5 keV \cite{Amaro:2022gub} could be further lowered by the increase in light yield down to almost 0.25 keV, opening the possibility to search for DM around 0.5 GeV/c$^2$ with the He target. At the same time, if more importance is given to the directional capabilities, such as HT recognition and angular resolution, the reduced diffusion of the \textit{Tt} can help improving the topology of the data for the imaging of the recoil tracks. For the above mentioned reasons, this study is deemed extremely important for the development of future recoiling imaging experiments.

\section{Conclusions}
In the context of a Dark Matter direct search, the CYGNO experiment is following an innovative path for directional detection to surpass the limitations of the current direct detection experiments. CYGNO is a gaseous TPC operated with a gas mixture of He:CF$_4$ mixture. The property of the gas allows the production of light during the amplification processes which is constituted by a stack of GEMs. The produced photons are readout optically with a combination of sCMOS cameras and PMTs. The optical readout  provides advantages as the sensors can be placed far from the amplification stage to reduce radioactivity in the sensitive volume and to image a large area with a single sensor. The amount of photons generated per secondary electrons in the gas and the solid angle covered by the sensor diminish the intensity of the signal detected and could pose limitation of the energy threshold of the experiment.
In order to improve the photon yield and minimise the intrinsic diffusion caused by the amplification stage itself, a study is performed to analyse the potential of improvement of different amplification structures and helium content in the gas mixture. Moreover, following previous studies, the introduction of a strong electric field below the outermost GEM is also taken into consideration for the light yield improvements. Two of the collaboration prototypes, MANGO and LEMOn, were utilised for the analyses. Different combination of GEMs were mounted ranging from the typical triple 50$\mu$m thick ($t$) to a combination of GEMs including a thicker 125 $\mu$m thick ($T$). The largest light output is achieved with the triple layer which also provides the best energy resolution, but returns a high diffusion that is also dependent on the voltage applied across the GEMs. Using only two GEMs results in a reduction of the diffusion and loss of light output and energy resolution. The smallest diffusion is obtained with a combination of one thick and one thin GEM. The standard 60/40 mixture of He:CF$_4$ is the one that allows the best stability in terms of spark occurrence. Increasing the amount of helium does not bring any sizable advantage  other than reduced voltage required to bias the GEMs and smaller density (good for tracking but less for exposure purposes).\\
The addition of the electric field in the induction region is analysed in depth. By measuring the relative increase in light and charge output with increasing induction field leads to the recognition of a region at high fields wherein the light increase surpasses the one of the charge. The Ansys Maxwell simulations of the electric fields suggest that the introduction of a strong induction field modifies the profile of the electric field of the GEM so that a small region few tens of $\mu$m below the GEM holes it becomes strong enough to allow the multiplication and the neutral fragmentation of CF$_4$. Since the strength of the electric field is not at the level of the one inside the GEM hole, the process favoured is the neutral fragmentation over the ionization, which results in a larger amount of photons than electrons. When the geometry of the GEM closest to the induction electrode is changed with a thicker GEM, the larger dimension of the hole and the weaker difference of field intensity between inside and outside, makes this effect more powerful causing a higher light enhancement with a slightly faster degradation of energy resolution and larger diffusion.\\
In terms of performances, the addition of this strong field below the GEM combined with a different amplification structure can lead a two stack of GEMs to yield an amount of light which recovers the one granted by a triple layer, but with a smaller diffusion from the amplification stage itself. Careful analyses of the light yield, diffusion and energy resolution permitted to discover that each type of structure combined with a strong induction field was found to excel in the measurement of a specific observable: a triple $t$ stack had the best energy resolution, a double $T$ one had the best light yield, and a $Tt$ one had the lowest intrinsic diffusion. Therefore, depending on the type of search and purpose, the best amplification structure can be employed to maximise performances. For example, the possibility of improving by a factor 2 the light yield of a CYGNO detector, coupled to the technological advancements of CMOS-based optical devices, could lead to a reduction of the energy threshold down to few hundreds of eV, firmly improving the sensitivity to below GeV/c$^2$ WIMP masses. Nonetheless, in applications where the topology of the recoiling track is more important than the light yield, a $Tt$ configuration could be foreseen in order to minimise diffusion. This innovative way of employing the GEM amplification structure in a He:CF$_4$ mixture results, therefore, extremely relevant for several optical TPC applications, even beyond dark matter searches.

\section*{Acknowledgement}
This project has received funding under the European Union’s Horizon 2020 research and innovation programme from the European Research Council (ERC) grant agreement No 818744.

\bibliographystyle{spphys}
\bibliography{Reference.bib}

\end{document}